\documentclass[10pt,onecolumn,aps,prd,preprintnumbers,showpacs,superscriptaddress,nofootinbib,amsmath,amssymb,floats,floatfix,showkeys,notitlepage,longbibliography]{revtex4-2} 
\usepackage{orcidlink}
\usepackage{comment}
\usepackage{lipsum}
\usepackage{graphicx}
\usepackage{subfigure}
\usepackage{palatino}
\usepackage{sans}
\usepackage{hyperref}
\hypersetup{colorlinks=true,linkcolor=blue,urlcolor=blue,citecolor=blue}
\usepackage[toc,page]{appendix}
\usepackage[normalem]{ulem}
\usepackage{adjustbox}
\usepackage{latexsym}
\usepackage{amsmath}
\numberwithin{equation}{section}
\usepackage{amssymb}
\usepackage{amsfonts}
\usepackage{dcolumn}
\usepackage{bm}
\usepackage{tikz}
\usetikzlibrary{decorations.pathmorphing}
\usepackage{bigints}
\usepackage{array,tabularx,multirow,booktabs}
\usepackage[tracking=true]{microtype}
\usepackage{soul} 
\SetTracking{}{500}
\SetTracking{encoding={*}, shape=sc}{40}
\UseRawInputEncoding 
\allowdisplaybreaks
\usepackage[utf8]{inputenc}
\usepackage{xcolor} 
\usepackage{lmodern}

\begin{document} \sloppy

\title{Gauss-Bonnet lensing of spinning massive particles in static spherically symmetric spacetimes}

\author{Reggie C. Pantig \orcidlink{0000-0002-3101-8591}} 
\email{rcpantig@mapua.edu.ph}
\affiliation{Physics Department, School of Foundational Studies and Education, Map\'ua University, 658 Muralla St., Intramuros, Manila 1002, Philippines.}

\author{Ali \"Ovg\"un \orcidlink{0000-0002-9889-342X}}
\email{ali.ovgun@emu.edu.tr}
\affiliation{Physics Department, Eastern Mediterranean University, Famagusta, 99628 North
Cyprus via Mersin 10, Turkiye.}

\begin{abstract}
We extend the finite-distance Jacobi-metric Gauss-Bonnet framework of Li \textit{et al}. [10.1103/PhysRevD.101.124058] [2006.13047] to massive test particles carrying intrinsic spin. At pole-dipole order, the Mathisson-Papapetrou-Dixon dynamics generically drives the spatial ray away from Jacobi geodesics, so the standard Gauss-Bonnet construction must be reformulated to accommodate a non-geodesic particle boundary. Working in the aligned-spin planar sector with the Tulczyjew-Dixon spin supplementary condition and retaining terms linear in the spin, we derive a spin-generalized deflection identity in which the spin dependence enters through a single additional boundary functional: the geodesic-curvature integral of the physical ray in the Jacobi manifold. We show that Li’s circular-orbit boundary choice remains fully compatible with this generalization and continues to collapse the Gaussian-curvature surface term to an effective one-dimensional integral. We then provide an implementation-ready weak-field recipe that relates the required geodesic curvature directly to the MPD spin-curvature force, enabling systematic perturbative evaluation without introducing model-dependent definitions of asymptotic angles. As applications, we validate the Schwarzschild limit, including the expected linear-in-spin weak-field scaling, and compute leading spin corrections for Reissner-Nordstr\"om and Kottler (Schwarzschild-de Sitter) geometries with finite source and receiver distances. In Kottler, we show that the constant-curvature part of the cosmological constant does not generate a linear-in-spin MPD force under the Tulczyjew–Dixon condition; nevertheless, the finite-distance spin correction acquires an explicit $\Lambda$-dependence through the Jacobi-metric prefactor entering the Gauss-Bonnet boundary functional, in addition to the Weyl-driven (mass-sourced) contribution
\end{abstract}


\maketitle

\section{Introduction}\label{sec:intro}

Gravitational deflection of particles by compact objects is a central observable in relativistic astrophysics and, more broadly, a clean probe of how spacetime geometry governs propagation. It spans cosmological applications such as CMB lensing \cite{Lewis:2006fu} and connects the classical tests of general relativity-from early successes in solar-system dynamics \cite{Einstein_2005,Weinberg:1972kfs,Will:2014kxa}-to horizon-scale observations of compact objects \cite{EventHorizonTelescope:2019dse,EventHorizonTelescope:2022wkp,Vagnozzi:2022moj}. In asymptotically flat spacetimes, the bending of null and timelike trajectories encodes detailed information about the exterior geometry, including the structure of photon surfaces and critical curves \cite{Claudel:2000yi,Bozza:2002zj,Tsukamoto:2016jzh}. The strong-deflection program and its image phenomenology build on an extensive analytic literature on relativistic lensing, magnifications, and time delays \cite{Virbhadra:1999nm,Virbhadra:2002ju,Virbhadra:2007kw,Virbhadra:2008ws,Virbhadra:2022iiy,Virbhadra:2024xpk}, with early extensions covering naked-singularity lensing and scalar-field effects \cite{Virbhadra:1998kd,Virbhadra:1998dy} and more recent studies quantifying how additional physics (e.g.\ a cosmological constant) reshapes photon-sphere and shadow radii \cite{Adler:2022qtb,Virbhadra:2022ybp}. These developments are now tightly linked to black-hole imaging, where photon rings and lensing rings provide precision targets for strong-field tests \cite{Gralla:2019xty,Johnson:2019ljv,Gralla:2020srx,Aratore:2024bro,Perlick:2021aok}.

A particularly influential perspective is the geometric approach initiated by Gibbons and Werner, in which the deflection of light in a static lens is related to the Gauss-Bonnet theorem (GBT) applied to a suitable two-dimensional Riemannian metric \cite{Gibbons:2008rj,Perlick:2004tq,Werner:2012rc,Perlick_1990}. This program has been extended to finite-distance configurations, where both source and receiver lie at finite radii, yielding an intrinsic and operationally meaningful definition of the deflection angle that remains applicable beyond asymptotically flat settings \cite{Ishihara:2016vdc,Ishihara:2016sfv,Ono:2017pie,Takizawa:2020egm,Ono:2019hkw}. On the observational side, the EHT images of M87* and Sagittarius~A* have elevated shadow and photon-ring morphology to primary diagnostics of compact-object spacetimes \cite{EventHorizonTelescope:2019dse,EventHorizonTelescope:2022wkp}, motivating broad theory efforts mapping beyond-Schwarzschild parameters into shadow and lensing observables \cite{Khodadi:2020jij,Allahyari:2019jqz,Vagnozzi:2022moj,Vincent:2022fwj,Chen:2022kzv}, including explicit links between strong lensing and ringdown observables in dynamical ``lensing tomography'' \cite{Zhong:2024ysg}. Complementary messengers-quasinormal-mode spectra and late-time tails-probe the same strong-field region \cite{Regge:1957td,Zerilli:1970wzz,Moncrief:1974am,Martel:2005ir,Price:1971fb,Kokkotas:1999bd,Berti:2009kk}, and the broader interplay between lensing and gravitational waves continues to develop, from wave-optics/lensed-waveform studies to the bending of light by gravitational waves \cite{Kaiser:1996wk,Damour:1998jm,Kopeikin:1999ev,Dai:2017huk}. In this context, time-dependent or wave-like perturbations of the gravitational field can imprint themselves on shadows and weak-field deflection \cite{Wang:2019skw,Pantig:2024kfn}, while spontaneous Lorentz-symmetry breaking can source primordial gravitational-wave signatures and modify the gravitational sector in ways that may ultimately be tested by precision lensing and imaging \cite{Khodadi:2025wuw}.

A realistic assessment of lensing observables must also account for propagation media and environmental effects. Plasma induces frequency-dependent refraction that can significantly modify bending angles and image properties near compact objects \cite{Morozova:2013uyv,Schee:2017hof,Turimov:2018ttf}, and strong/retro-lensing has been explored in a variety of nonstandard backgrounds, including braneworld black holes \cite{Abdujabbarov:2017pfw}. Lensing has likewise been computed for effective or quantum-corrected geometries \cite{Turakhonov:2024xfg,Wang:2025fmz,Ditta:2025vsa} and for astrophysically ``dressed'' lenses such as charged black holes embedded in dark-matter halos \cite{Molla:2025yoh}. Related developments include neutrino lensing in parametrized spacetimes \cite{Alloqulov:2024sns} and lensing/shadow studies of nonlinear-electrodynamics black holes, including ModMax and magnetically charged solutions \cite{Pantig:2022gih,Allahyari:2019jqz}. These directions sharpen the need for lensing frameworks that are geometrically transparent, adaptable to finite-distance configurations, and robust under controlled departures from vacuum geodesic propagation.

For \emph{massive} particles, an efficient adaptation of the GBT program uses the Jacobi metric \cite{Gibbons:2015qja,Crisnejo:2018uyn,Crisnejo:2019ril,Zhang:2022rnn}: at fixed conserved energy, the spatial projection of a timelike geodesic can be represented as a geodesic of an energy-dependent Riemannian metric on the spatial manifold, allowing one to import the GBT method from null to timelike motion. Li \textit{et al}. developed a particularly useful implementation in static spherically symmetric (SSS) spacetimes by introducing an auxiliary circular-orbit boundary that collapses the GBT surface integral into a one-dimensional expression while preserving finite-distance corrections \cite{Gibbons:2015qja,Li:2020wvn,Li:2019qyb}. Such constructions dovetail naturally with the classical geometry of compact-object spacetimes \cite{Schwarzschild:1916uq,Chandrasekhar:1985kt,Cunningham_1972,Capozziello:2025wwl,DeBianchi:2025bgn,Capozziello:2024ucm}.

The goal of the present work is to synthesize this Jacobi-Gauss-Bonnet framework with the dynamics of a \emph{spinning} massive test particle \cite{Wald:1972sz,Hanson:1974qy,Bailey:1975fe}. At pole-dipole order, the motion of a spinning body is governed by the Mathisson-Papapetrou-Dixon (MPD) equations, which describe spin-curvature coupling and require a spin supplementary condition (SSC) to fix the representative worldline of the extended body \cite{Papapetrou:1951pa,Corinaldesi:1951pb,Dixon:1970zza,Dixon:1970zz,Dixon:1974xoz,Ehlers:1977gyn,Lukes-Gerakopoulos:2014dma}. A key obstruction then appears: even in a static background, spin-curvature coupling generically drives the particle away from geodesic motion, and the spatial ray is therefore not a geodesic of the Jacobi geometry that underlies the Li-type construction. This non-geodesic character is the defining physical feature of spin-induced corrections and must be incorporated into the geometric method in an invariant way; general discussions of SSC dependence and related subtleties in MPD dynamics underscore the importance of making this step explicit \cite{Lukes-Gerakopoulos:2017cru}.

We address this by formulating a Gauss-Bonnet lens-domain construction in which the physical spinning-particle ray is permitted to be non-geodesic in the Jacobi manifold, while the auxiliary circular boundary is retained as a Jacobi geodesic in the sense used by Li \textit{et al}. The resulting framework keeps the finite-distance deflection-angle definition intact and preserves the circular-orbit simplification of the surface term, while isolating spin effects into a boundary functional that measures the intrinsic non-geodesicity of the ray. We then develop a weak-field evaluation strategy that makes this boundary functional computable directly from MPD input at linear order in the spin, enabling systematic applications to standard SSS black-hole families. We illustrate the method with charged and $\Lambda$-dressed lenses, emphasizing connections to existing observational programs and to the broader landscape of strong-field tests enabled by lensing, shadows, and ringdowns \cite{Perlick:2021aok,Vagnozzi:2022moj,Will:2014kxa}.

The paper is organized as follows. In Section \ref{sec2} we review the geometry and notation of Li \textit{et al}., emphasizing the Jacobi metric for massive particles, the circular-orbit boundary condition, and the finite-distance Gauss-Bonnet definition of the deflection angle. Section \ref{sec3} introduces the pole-dipole MPD dynamics with our SSC choice and derives the effective first integrals needed for SSS scattering with aligned spin. In Section \ref{sec4} we present the Gauss-Bonnet construction adapted to a non-geodesic particle ray and define the corresponding deflection-angle functional. Section \ref{sec5} turns the new spin-dependent term into an implementation-ready weak-field recipe by relating the geodesic curvature of the ray to the MPD spin-curvature force and by specifying a consistent perturbative expansion scheme. Section \ref{sec6} applies the formalism to Reissner-Nordstr\"om and Kottler spacetimes, using Schwarzschild as a validation benchmark. Section \ref{sec7} collects consistency checks and physical interpretation, and Section \ref{sec8} concludes.

We work in units $G=c=1$ and adopt the metric signature $(-,+,+,+)$.  

\section{Review of Li \textit{et al}.'s geometry and notation} \label{sec2}
In this section we review the geometric framework and notation introduced by Li \textit{et al}. for treating weak gravitational deflection of massive particles using a Jacobi-metric formulation on a two-dimensional Riemannian manifold. The essential point is that, for a static spherically symmetric (SSS) spacetime, timelike geodesic motion of fixed conserved energy can be re-expressed as geodesic motion in an energy-dependent Riemannian metric on a spatial slice. This reformulation is the backbone of the Gauss-Bonnet approach, because it allows us to define Gaussian curvature intrinsically on the two-dimensional lens plane and to characterize special boundary curves, including circular orbits, by their geodesic curvature.

\subsection{SSS and Jacobi metric for massive particles} \label{sec2.1}
Let $(t,r,\theta\,\phi)$ be standard spherical coordinates, and consider a generic static spherically symmetric spacetime of the form \cite{Perlick:2004tq,Perlick_1990}
\begin{equation}
ds^2 = -A(r)\,dt^2 + B(r)\,dr^2 + C(r)\Big(d\theta^2+\sin^2\theta\,d\phi^2\Big),
\label{2.1}
\end{equation}
where $A(r)>0$, $B(r)>0$, and $C(r)>0$ in the domain of outer communication. For asymptotically flat spacetimes one has $A(r)\to 1$, $B(r)\to 1$, and $C(r)\sim r^2$ as $r\to\infty$.
Let $m$ denote the particle rest mass. We take $p^\mu=m u^\mu$ with
$u^\mu\equiv dx^\mu/d\tau$ and $g_{\mu\nu}u^\mu u^\nu=-1$; covariant components are
$p_\mu\equiv g_{\mu\nu}p^\nu=m u_\mu$. Stationarity and spherical symmetry imply two conserved quantities along timelike geodesics: the energy $E\equiv -p_t$ and the azimuthal angular momentum $L\equiv p_\phi$. Restricting without loss of generality to the equatorial plane $\theta=\pi/2$ with $u^\theta=0$, these constants read
\begin{equation}
E = m\,A(r)\,\frac{dt}{d\tau},
\qquad
L = m\,C(r)\,\frac{d\phi}{d\tau}.
\label{2.2}
\end{equation}
Following the Jacobi-metric construction for static spacetimes (and the usage adopted by Li \textit{et al}.), we introduce an energy-dependent Riemannian metric on the spatial slice $t=\mathrm{const}$ such that the spatial projections of timelike geodesics of fixed $E$ coincide with geodesics of this Riemannian metric. Concretely, define the Jacobi line element $d\ell_J^2$ by \cite{Gibbons:2015qja,Zhang:2022rnn}
\begin{equation}
d\ell_J^2=
\frac{E^2-m^2A(r)}{A(r)}
\left[
B(r)\,dr^2 + C(r)\Big(d\theta^2+\sin^2\theta\,d\phi^2\Big)
\right].
\label{2.3}
\end{equation}
The conformal prefactor is positive whenever $E^2>m^2A(r)$, ensuring that the induced Jacobi geometry is Riemannian in the region accessible to the particle. On the equatorial plane $\theta=\pi/2$ this reduces to the two-dimensional Jacobi metric
\begin{equation}
d\ell_J^2=
\frac{E^2-m^2A(r)}{A(r)}
\left[
B(r)\,dr^2 + C(r)\,d\phi^2
\right].
\label{2.4}
\end{equation}
For an asymptotically flat spacetime, $E$ admits the standard interpretation as the energy measured at infinity. Writing $v$ for the particle speed at infinity as measured by asymptotic static observers, we parameterize
\begin{equation}
E = m \gamma, \qquad \gamma \equiv \left(1 - v^2\right)^{-1/2}
\label{2.5}
\end{equation}
It is also convenient to introduce the (asymptotic) impact parameter $b$ by matching the asymptotic angular momentum to its flat-space form. Since the spatial momentum at infinity satisfies $p_\infty=\sqrt{E^2-m^2}=Ev$, we define
\begin{equation}
L = b\,\sqrt{E^2-m^2} = b\,E\,v.
\label{2.6}
\end{equation}
With these definitions, the orbit equation in the equatorial plane follows directly from the normalization condition and the constants of motion. Eliminating $\tau$ in favor of $\phi$, we obtain
\begin{equation}
\left(\frac{dr}{d\phi}\right)^2=
\frac{C(r)}{B(r)}
\left[
\frac{C(r)}{b^2}\,
\frac{E^2-m^2A(r)}{A(r)\,(E^2-m^2)}
-1
\right].
\label{2.7}
\end{equation}
The massless limit is recovered by setting $m\to 0$ while keeping $E$ finite; in this limit the Jacobi metric becomes proportional to the usual optical metric, and Eq. \eqref{2.7} reduces to the standard null orbit equation with $b=L/E$.

\subsection{Circular orbit boundary and the \texorpdfstring{$\kappa(\gamma_{co})=0$}{} condition} \label{sec2.2}
A distinctive element in Li \textit{et al}.’s construction is the use of a circular orbit as a geometrically preferred boundary curve in the two-dimensional Jacobi manifold. Let $\gamma_{co}$ denote the coordinate circle $r=r_{co}$ in the equatorial plane, viewed as a curve embedded in the Riemannian manifold $(\mathcal{M}J,g^{(J)}_{ab})$ with line element \eqref{2.4}. Writing the Jacobi metric components as
\begin{equation}
g^{(J)}_{rr}(r)=\frac{E^2-m^2A(r)}{A(r)}\,B(r),
\qquad
g^{(J)}_{\phi\phi}(r)=\frac{E^2-m^2A(r)}{A(r)}\,C(r),
\label{2.8}
\end{equation}
the (signed) geodesic curvature of the circle $r=\mathrm{const}$ can be expressed purely in terms of these components. A convenient form is
\begin{equation}
\kappa(r)=
\frac{1}{2\sqrt{g^{(J)}_{rr}(r)}}\,
\frac{d}{dr}\ln g^{(J)}_{\phi\phi}(r)\,
\label{2.9}
\end{equation}
evaluated at the chosen radius. The defining condition for $\gamma_{co}$ to be a geodesic of the Jacobi geometry is
\begin{equation}
\kappa(\gamma_{co})=0,
\label{2.10}
\end{equation}
which is equivalent to the vanishing of the radial derivative of $g^{(J)}_{\phi\phi}$. Using Eq. \eqref{2.8}, this yields the compact circular-orbit condition
\begin{equation}
\frac{d}{dr}
\left[
\frac{E^2-m^2A(r)}{A(r)}\,C(r)
\right]_{r=r_{co}}
=0.
\label{2.11}
\end{equation}
Notably, the function $B(r)$ drops out of Eq. \eqref{2.11}, so the location of the Jacobi-geodesic circular orbit is determined entirely by $A(r)$ and $C(r)$ together with the particle parameters $(E,m)$, exactly as emphasized by Li \textit{et al}. Expanding Eq. \eqref{2.11} gives an equivalent logarithmic form,
\begin{equation}
\left.\frac{d}{dr}\ln C(r)\right|_{r=r_{co}}-
\left.\frac{d}{dr}\ln A(r)\right|_{r=r_{co}}-
\left.\frac{m^2 A'(r)}{E^2-m^2A(r)}\right|_{r=r_{co}}
=0,
\label{2.12}
\end{equation}
where a prime denotes differentiation with respect to $r$. In the null limit $m\to 0$, Eq. \eqref{2.12} reduces to $d\ln(C/A)/dr=0$, i.e. $(C/A)'=0$, reproducing the familiar photon-sphere condition for SSS spacetimes. For timelike motion, Eq. \eqref{2.12} encodes the energy dependence of the circular-orbit radius: increasing $E/m$ (equivalently increasing the asymptotic speed $v$) continuously interpolates between bound timelike circular orbits and the null circular orbit.
Finally, we stress the geometric role of Eq. \eqref{2.10}: choosing $\gamma_{co}$ to satisfy $\kappa(\gamma_{co})=0$ ensures that this circular boundary contributes no geodesic-curvature term when applying the Gauss-Bonnet theorem to a lensing domain whose boundary includes $\gamma_{co}$. This simplification is central to Li \textit{et al}.’s treatment of weak deflection in settings where one prefers a finite integration domain.

\subsection{Gauss-Bonnet setup and deflection-angle definition} \label{sec2.3}
We denote by $(\mathcal{M},g_{ij})$ the two-dimensional Riemannian manifold obtained by restricting the Jacobi metric to the equatorial plane, with line element given in Eq. \eqref{2.4}. Let $D\subset\mathcal{M}$ be a compact, oriented domain with piecewise smooth boundary $\partial D$, Gaussian curvature $K$ associated with $g_{ij}$, and Euler characteristic $\chi(D)$. The Gauss-Bonnet theorem states that
\begin{equation}
\iint_{D}K\,dS+\oint_{\partial D}\kappa_g\,d\sigma+\sum_{i=1}^{N}\beta_i=2\pi\chi(D),
\label{2.13}
\end{equation}
where $dS=\sqrt{\det g}\,dr\,d\phi$ is the area element in local coordinates $(r,\phi)$, $d\sigma$ is the arclength element along $\partial D$, $\kappa_g$ is the geodesic curvature of the boundary (with respect to the inward-pointing normal), and $\beta_i$ are the exterior jump angles at the $N$ vertices of $\partial D$.
Following Li \textit{et al}., we define the finite-distance deflection angle by the Ishihara-type prescription \cite{Ishihara:2016vdc}
\begin{equation}
\alpha\equiv \Psi_R-\Psi_S+\phi_{RS},
\label{2.14}
\end{equation}
where $\phi_{RS}\equiv \phi_R-\phi_S$ is the coordinate angular separation between receiver and source, while $\Psi_R$ and $\Psi_S$ are the angles between the tangent direction of the particle trajectory and the outward radial direction at the receiver $R$ and source $S$, respectively, all defined intrinsically in the Jacobi geometry on $\mathcal{M}$.
The lensing domain $D$ used by Li \textit{et al}. is bounded by four curves in $\mathcal{M}$: (i) the spatial projection of the particle trajectory, denoted $\gamma_g$, connecting $S$ to $R$; (ii) an auxiliary circular arc $\gamma_{co}$ at $r=r_{co}$; and (iii) two outgoing radial curves at fixed $\phi=\phi_S$ and $\phi=\phi_R$ joining $S$ and $R$ to $\gamma_{co}$. For the diagonal, axisymmetric Jacobi metric \eqref{2.4}, these radial curves are Jacobi geodesics: since $\partial_\phi g_{ij}=0$ and $\Gamma^{\phi}{}_{rr}=0$, curves with $\phi=\mathrm{const}$ satisfy the Jacobi geodesic equation. In the spinless setting reviewed here, $\gamma_g$ is a geodesic of $(\mathcal{M},g_{ij})$ by construction of the Jacobi metric, and the radial curves are also geodesics in the axisymmetric diagonal metric \eqref{2.4}. The circular arc $\gamma_{co}$ is chosen to satisfy the Jacobi-geodesic condition $\kappa(\gamma_{co})=0$ discussed in Section \ref{sec2.2}, so that it is likewise a geodesic in $\mathcal{M}$. With this choice, the boundary geodesic-curvature term in Eq. \eqref{2.13} vanishes:
\begin{equation}
\oint_{\partial D}\kappa_g\,d\sigma=0.
\label{2.15}
\end{equation}
The boundary $\partial D$ has four vertices: at $S$ and $R$, where $\gamma_g$ meets the radial curves, and at the two intersection points where the radial curves meet $\gamma_{co}$. Since $\gamma_{co}$ is a coordinate circle and the radial curves are coordinate radials, they intersect orthogonally in the Jacobi geometry, contributing two right angles whose sum is $\pi$. The domain $D$ is nonsingular, so $\chi(D)=1$. Substituting these facts into Eq. \eqref{2.13} yields
\begin{equation}
\iint_{D}K\,dS+\beta_S+\beta_R=\pi,
\label{2.16}
\end{equation}
where $\beta_S$ and $\beta_R$ are the exterior jump angles at $S$ and $R$, respectively. With the orientation conventions used by Li \textit{et al}., these are related to $\Psi_S$ and $\Psi_R$ by $\beta_S=\Psi_S$ and $\beta_R=\pi-\Psi_R$.
Equation \eqref{2.16} then becomes $\iint_D K\,dS=\Psi_R-\Psi_S$, and inserting this result into Eq. \eqref{2.14} gives the central spinless deflection formula
\begin{equation}
\alpha=\iint_{D}K\,dS+\phi_{RS}.
\label{2.17}
\end{equation}
This expression remains meaningful when $S$ and $R$ are at finite distances, and, crucially for Li \textit{et al}.’s applications, it avoids any reference to asymptotic regions that may not exist in asymptotically non-flat geometries.

\subsection{Circular-orbit simplification of the surface integral} \label{sec2.4}
We next summarize the technical simplification emphasized by Li \textit{et al}.: when the lower radial boundary of $D$ is a Jacobi-geodesic circular orbit $\gamma_{co}$, the surface term in Eq. \eqref{2.17} can be reduced to a one-dimensional integral that only requires evaluating a radial primitive at the physical trajectory $r=r(\phi)$.
Let $g_{ij}$ be the Jacobi metric on the equatorial plane in coordinates $(r,\phi)$, and denote its determinant by $\det g$. A useful expression for the Gaussian curvature in two dimensions is
\begin{equation}
K=\frac{1}{\sqrt{\det g}}
\left[
\frac{\partial}{\partial \phi}\!\left(\frac{\sqrt{\det g}}{g_{rr}}\Gamma^{\phi}{}_{rr}\right)-
\frac{\partial}{\partial r}\!\left(\frac{\sqrt{\det g}}{g_{rr}}\Gamma^{\phi}{}_{r\phi}\right)
\right],
\label{2.18}
\end{equation}
where $\Gamma^{i}{}_{jk}$ are the Christoffel symbols of $g_{ij}$.
For the SSS Jacobi metric, the components depend only on $r$, so the $\phi$-derivative term vanishes and Eq. \eqref{2.18} reduces to a total radial derivative,
\begin{equation}
K\sqrt{\det g}=-\frac{d}{dr}\left(\frac{\sqrt{\det g}}{g_{rr}}\Gamma^{\phi}{}_{r\phi}\right).
\label{2.19}
\end{equation}
It is therefore natural to introduce the radial primitive (choosing the irrelevant additive constant so that the following equality holds),
\begin{equation}
\mathcal{I}(r)\equiv \int^{r}_{r_{co}} K\sqrt{\det g}\,dr'
=-\frac{\sqrt{\det g}}{g_{rr}}\Gamma^{\phi}{}_{r\phi}.
\label{2.20}
\end{equation}
For a diagonal axisymmetric metric, $\Gamma^{\phi}{}_{r\phi}=\frac{1}{2}\,d(\ln g_{\phi\phi})/dr$, so Eq. \eqref{2.20} can be written more explicitly as
\begin{equation}
\mathcal{I}(r)=-\frac{1}{2}\sqrt{\frac{g_{\phi\phi}(r)}{g_{rr}(r)}}\,\frac{d}{dr}\ln g_{\phi\phi}(r).
\label{2.21}
\end{equation}
Comparing Eq. \eqref{2.21} with the geodesic curvature of a coordinate circle $r=\mathrm{const}$ in $(\mathcal{M},g_{ij})$, which was written in Eq. \eqref{2.9} as $\kappa(r)=\frac{1}{2\sqrt{g_{rr}}}\,d(\ln g_{\phi\phi})/dr$, we obtain the identity
\begin{equation}
\mathcal{I}(r)=-\sqrt{g_{\phi\phi}(r)}\,\kappa(r).
\label{2.22}
\end{equation}
In particular, when the circular arc $\gamma_{co}$ is chosen to satisfy $\kappa(\gamma_{co})=0$, we have
\begin{equation}
\mathcal{I}(r_{co})=0,
\label{2.23}
\end{equation}
where we have used the freedom to shift $\mathcal{I}(r)$ by an additive constant, which cancels in the definite integral below.
Let $r=r(\phi)$ denote the radial coordinate of the particle trajectory $\gamma_g$ in the Jacobi geometry. The domain $D$ is then described by $\phi\in[\phi_S,\phi_R]$ and $r\in[r_{co},r(\phi)]$. Using $dS=\sqrt{\det g}\,dr\,d\phi$ and Eq. \eqref{2.20}, the curvature surface term becomes
\begin{equation}
\iint_{D}K\,dS=
\int_{\phi_S}^{\phi_R}\int_{r_{co}}^{r(\phi)}K\sqrt{\det g}\,dr\,d\phi=
\int_{\phi_S}^{\phi_R}\Big[\mathcal{I}(r(\phi))-\mathcal{I}(r_{co})\Big]d\phi.
\label{2.24}
\end{equation}
Invoking Eq. \eqref{2.23} yields the simplified form
\begin{equation}
\iint_{D}K\,dS=
\int_{\phi_S}^{\phi_R}\mathcal{I}(r(\phi))\,d\phi\,
\label{2.25}
\end{equation}
and substituting this into Eq. \eqref{2.17} gives the practical Li \textit{et al}. representation of the finite-distance deflection angle,
\begin{equation}
\alpha=\int_{\phi_S}^{\phi_R}\mathcal{I}(r(\phi))\,d\phi+\phi_{RS}.
\label{2.26}
\end{equation}
The key advantage of Eq. \eqref{2.26} is that the circular-orbit boundary removes any explicit lower-limit contribution from $r=r_{co}$, thereby reducing the evaluation of the Gauss-Bonnet surface term to a single primitive evaluated along the physical trajectory.

\section{Spinning massive particle dynamics in SSS: working assumptions and constants} \label{sec3}
We now generalize the spinless setup reviewed in Section \ref{sec2} to a massive particle carrying intrinsic spin. The appropriate test-body description is provided by the Mathisson-Papapetrou-Dixon (MPD) equations at pole-dipole order, which capture spin-curvature coupling while neglecting higher multipole moments. In this framework the spatial trajectory is generically non-geodesic in the Jacobi geometry, which is precisely the feature that will ultimately necessitate a Gauss-Bonnet treatment accommodating nonzero geodesic curvature of the physical path.

\subsection{MPD equations at pole-dipole order and SSC choice} \label{sec3.1}
We consider a fixed background spacetime with metric $g_{\mu\nu}$ of signature $(-,+,+,+)$ and Levi-Civita connection $\nabla_\mu$. A representative worldline $z^\mu(\tau)$ is parametrized by proper time $\tau$, with tangent (four-velocity) defined by $u^\mu\equiv dz^\mu/d\tau$ and normalized as $u^\mu u_\mu=-1$. The covariant derivative along the worldline is denoted by
\begin{equation}
\frac{D}{d\tau}\equiv u^\alpha\nabla_\alpha.
\label{3.1}
\end{equation}
At pole-dipole order, the dynamical variables are the momentum $p^\mu(\tau)$ and the antisymmetric spin tensor $S^{\mu\nu}(\tau)=-S^{\nu\mu}(\tau)$. The MPD equations governing their evolution are \cite{Papapetrou:1951pa,Dixon:1970zza,Wald:1972sz}
\begin{equation}
\frac{D p^\mu}{d\tau}=-\frac{1}{2}\,R^\mu{}_{\nu\alpha\beta}\,u^\nu S^{\alpha\beta},
\label{3.2}
\end{equation}
\begin{equation}
\frac{D S^{\mu\nu}}{d\tau}=p^\mu u^\nu-p^\nu u^\mu,
\label{3.3}
\end{equation}
where $R^\mu{}_{\nu\alpha\beta}$ is the Riemann tensor of $g_{\mu\nu}$. We adopt the curvature convention
\begin{equation}
R^\rho{}{\sigma_{\mu\nu}}=\partial_\mu\Gamma^\rho_{\nu\sigma}-\partial_\nu\Gamma^\rho_{\mu\sigma}
+\Gamma^\rho_{\mu\lambda}\Gamma^\lambda_{\nu\sigma}-\Gamma^\rho_{\nu\lambda}\Gamma^\lambda_{\mu\sigma},
\label{3.4}
\end{equation}
with $\Gamma^\rho_{\mu\nu}$ the Christoffel symbols, and $R_{\mu\nu}=R^\rho{}_{\mu\rho\nu}$.
Equations \eqref{3.2} and \eqref{3.3} alone do not uniquely determine the worldline because the notion of center of mass is gauge-like for an extended body in relativity. One must impose a spin supplementary condition (SSC) to select a representative centroid and close the system \cite{Dixon:1970zza}. In this work we adopt the Tulczyjew-Dixon SSC,
\begin{equation}
p_\mu S^{\mu\nu}=0,
\label{3.5}
\end{equation}
which defines the centroid in the rest frame of the momentum. This choice is standard in relativistic test-body dynamics and is particularly convenient because it yields two Casimir-type invariants that remain constant along the motion at pole-dipole order \cite{Dixon:1974xoz,Ehlers:1977gyn}.
First, we define the invariant (dynamical) mass by
\begin{equation}
m^2\equiv -p_\mu p^\mu,
\label{3.6}
\end{equation}
which reduces to the usual rest mass in the spinless limit where $p^\mu$ is proportional to $u^\mu$. Under the Tulczyjew-Dixon SSC, $m$ is conserved along the trajectory for pole-dipole motion in a fixed background \cite{Dixon:1974xoz,Blanco:2023jxf}.
Second, we define the spin magnitude by
\begin{equation}
s^2\equiv \frac{1}{2}\,S_{\mu\nu}S^{\mu\nu},
\label{3.7}
\end{equation}
which is likewise conserved under Eq. \eqref{3.5} at pole-dipole order \cite{Dixon:1974xoz,Blanco:2023jxf}.
A key physical feature of the MPD system is that the momentum and four-velocity are not generally parallel when spin-curvature coupling is present. It is therefore useful to introduce the unit timelike momentum direction
\begin{equation}
v^\mu\equiv \frac{p^\mu}{m}\,
\qquad
v^\mu v_\mu=-1,
\label{3.8}
\end{equation}
so that the SSC \eqref{3.5} can be read as $v_\mu S^{\mu\nu}=0$, i.e. the spin tensor is purely spatial in the momentum rest frame. The deviation between $u^\mu$ and $v^\mu$ is responsible for the non-geodesic character of the spatial trajectory that will enter the Gauss-Bonnet construction once we project the motion onto the equatorial Jacobi manifold.
Finally, we state the regime of validity assumed throughout our analysis. The pole-dipole truncation treats the body as a test particle with intrinsic spin but neglects quadrupole and higher multipole moments. This approximation is controlled when the characteristic size $\ell$ of the body is much smaller than the local curvature radius $\mathcal{R}$ of the background, and when the intrinsic spin scale $s\sim m\ell$ satisfies $s/(m\mathcal{R})\ll 1$. Under these assumptions the MPD force term on the right-hand side of Eq. \eqref{3.2} consistently captures the leading spin-curvature interaction while higher-multipole corrections remain parametrically suppressed \cite{Dixon:1974xoz,Ehlers:1977gyn}.

\subsection{Planar reduction and spin alignment} \label{sec3.2}
We now specialize the pole-dipole dynamics to the symmetry sector relevant for lensing: motion confined to a two-dimensional plane with the particle spin aligned (or anti-aligned) with the orbital angular momentum. In the spinless case, planarity follows directly from spherical symmetry. For spinning bodies, generic initial data can excite out-of-plane motion through spin-curvature coupling; nevertheless, the aligned-spin sector defines a consistent invariant submanifold of the MPD system in static spherically symmetric backgrounds and is the standard setting adopted in analytic treatments of spinning-particle scattering and bound orbits \cite{Semerak:1999qc,Witzany:2023bmq}.

A central structural fact is that the MPD equations admit symmetry-generated first integrals. Let $\xi^\mu$ be any Killing vector of the background spacetime, satisfying $\nabla_{(\mu}\xi_{\nu)}=0$. For pole-dipole motion with the Tulczyjew-Dixon SSC, the combination
\begin{equation}
C_\xi \equiv \xi_\mu p^\mu+\frac{1}{2}\,\nabla_\mu\xi_\nu S^{\mu\nu}
\label{3.9}
\end{equation}
is conserved along the worldline \cite{Compere:2021kjz}.

In a static spherically symmetric spacetime of the form in Eq. \eqref{2.1}, there are four independent Killing vectors relevant here: the time-translation generator $\xi_{(t)}^\mu=(\partial_t)^\mu$ and the three rotational generators of $\mathrm{SO}(3)$. We define the conserved energy and the conserved azimuthal component of total angular momentum by
\begin{equation}
\mathcal{E}\equiv -C_{\xi_{(t)}},
\qquad
\mathcal{J}\equiv C_{\xi_{(\phi)}},
\label{3.10}
\end{equation}
where $\xi_{(\phi)}^\mu=(\partial_\phi)^\mu$. The signs are chosen so that $\mathcal{E}>0$ for future-directed motion and $\mathcal{J}$ reduces to the usual orbital angular momentum in the spinless limit.
The invariants associated with the three rotational Killing vectors may be viewed as the components of a conserved total angular momentum vector, including both orbital and spin contributions through the second term in Eq. \eqref{3.9}. Exploiting spherical symmetry, we may choose spatial axes such that this conserved total angular momentum points along the $z$-axis of our coordinate system. With this choice, the motion is confined to the equatorial plane $\theta=\pi/2$ provided the spin is aligned or anti-aligned with the same axis. Concretely, we impose the planar initial data
\begin{equation}
\theta=\frac{\pi}{2},
\qquad
p_\theta=0,
\qquad
u^\theta=0,
\label{3.11}
\end{equation}
together with an aligned-spin condition specified below. Reflection symmetry of the SSS geometry across the equatorial plane then ensures that the MPD evolution preserves Eq. \eqref{3.11} within the aligned-spin sector \cite{Witzany:2023bmq,Semerak:1999qc}.

To implement spin alignment in a covariant manner under the Tulczyjew-Dixon SSC, it is convenient to trade the antisymmetric spin tensor for a spin four-vector defined in the momentum rest frame. We introduce the Levi-Civita tensor with upper indices, normalized by
$\epsilon^{tr\theta\phi}=+1/\sqrt{-g}$ in the coordinate basis, where
$g\equiv \det(g_{\mu\nu})$, and define
\begin{equation}
s^{\mu}\equiv -\frac{1}{2m}\,\epsilon^{\mu\nu\alpha\beta}\,p_{\nu}S_{\alpha\beta}.
\label{3.12}
\end{equation}
By construction and the SSC \eqref{3.5}, $s^\mu p_\mu=0$, and the invariant spin magnitude introduced in Eq. \eqref{3.7} satisfies $s^2=s_\mu s^\mu$. Conversely, whenever $s^\mu p_\mu=0$, the spin tensor can be reconstructed as
\begin{equation}
S^{\mu\nu}=\frac{1}{m}\,\epsilon^{\mu\nu\alpha\beta}\,p_{\alpha}s_{\beta}.
\label{3.13}
\end{equation}
Equations \eqref{3.12} and \eqref{3.13} make explicit that spin degrees of freedom are purely spatial in the momentum rest frame and provide a clean parametrization of aligned-spin configurations \cite{Steinhoff:2012rw,Compere:2021kjz}.

In the equatorial reduction, we define aligned and anti-aligned spin by requiring that $s^\mu$ be orthogonal to the orbital plane and point along $\pm\partial_\theta$ at $\theta=\pi/2$. 
Equivalently, at $\theta=\pi/2$ we impose that the only nonvanishing component
of $s^{\mu}$ is $s^{\theta}$, with
\begin{equation}
s^{\theta}=\frac{\sigma\,s}{\sqrt{g_{\theta\theta}}}\bigg|_{\theta=\pi/2},
\qquad \sigma=\pm 1,
\label{3.14}
\end{equation}
where $\sigma=+1$ corresponds to alignment and $\sigma=-1$ to anti-alignment with the conserved total angular momentum direction fixed above. Under this condition, Eq. \eqref{3.13} implies that the only nonzero components of $S^{\mu\nu}$ lie in the $(t,r,\phi)$ subspace and, in particular, $S^{\mu\theta}=0$ on the equatorial plane. This aligned-spin restriction is the sector in which analytic SSS spinning-particle solutions and scattering observables are most naturally developed, and it is the sector we will adopt for the lensing problem.

\subsection{Effective first integrals and orbit equation with spin} \label{sec3.3}
We now make the conserved quantities introduced in Section \ref{sec3.2} explicit for the static spherically symmetric geometry \eqref{2.1} and for the aligned-spin equatorial sector specified in Section \ref{sec3.2}. We then combine these first integrals with the mass-shell constraint \eqref{3.6} to obtain a closed radial equation and an orbit equation for $r(\phi)$ including spin effects.
For pole-dipole motion with the Tulczyjew-Dixon SSC \eqref{3.5}, any Killing vector $\xi^\mu$ generates a conserved scalar $C_\xi$ defined in Eq. \eqref{3.9}. This is a standard consequence of Dixon’s formalism and is widely used in analytic treatments of spinning test bodies \cite{Dixon:1974xoz,Semerak:1999qc,Li:2020wvn}.
In particular, for the time-translation Killing vector $\xi_{(t)}^\mu=(\partial_t)^\mu$ and the axial Killing vector $\xi_{(\phi)}^\mu=(\partial_\phi)^\mu$, we define the constants of motion $\mathcal{E}$ and $\mathcal{J}$ as in Eq. \eqref{3.10}. To evaluate the associated spin couplings, we note that in the metric \eqref{2.1} the only nonvanishing derivatives of the corresponding covectors are radial. Writing a prime to denote $d/dr$, one has on the equatorial plane $\theta=\pi/2$
\begin{equation}
\nabla_r \xi^{(t)}{t}=\frac{1}{2}\,g_{tt}'=-\frac{1}{2}\,A'(r)\,
\qquad
\nabla_r \xi^{(\phi)}{\phi}=\frac{1}{2}\,g_{\phi\phi}'=\frac{1}{2}\,C'(r)\,
\label{3.15}
\end{equation}
with all other independent components vanishing or fixed by antisymmetry of $\nabla_\mu\xi_\nu$ for a Killing field.
Next, we specialize the spin tensor to the aligned-spin sector. Using the spin four-vector $s^\mu$ defined in Eq. \eqref{3.12}, the reconstruction formula \eqref{3.13}, and the aligned-spin condition \eqref{3.14}, one finds that the only independent nonzero components of $S^{\mu\nu}$ are $S^{rt}$, $S^{r\phi}$, and $S^{t\phi}$, which can be expressed directly in terms of the momentum components. It is convenient to introduce the (signed) spin-coupling function
\begin{equation}
\lambda(r)\equiv \frac{\sigma\,s}{m\,\sqrt{A(r)\,B(r)\,C(r)}},
\qquad
\sigma=\pm 1,
\label{3.16}
\end{equation}
so that the aligned-spin relations take the compact form
\begin{equation}
S^{rt}=\lambda(r)\,p_{\phi},\qquad 
S^{r\phi}=-\lambda(r)\,p_{t},\qquad 
S^{t\phi}=\lambda(r)\,p_{r},
\label{3.17}
\end{equation}
where $p_t=g_{tt}p^t$, $p_r=g_{rr}p^r$, and $p_\phi=g_{\phi\phi}p^\phi$.
Substituting Eqs. \eqref{3.15} and \eqref{3.17} into the general conserved charge \eqref{3.9} yields the explicit first-integral relations
\begin{equation}
\mathcal{E}=-p_t + \frac{1}{2}A'(r)\,\lambda(r)\,p_\phi,
\label{3.18}
\end{equation}
\begin{equation}
\mathcal{J}=p_\phi - \frac{1}{2}C'(r)\,\lambda(r)\,p_t.
\label{3.19}
\end{equation}
These equations reduce to the familiar spinless identifications $\mathcal{E}=-p_t$ and $\mathcal{J}=p_\phi$ when $s\to 0$. Moreover, for asymptotically flat spacetimes, $A'(r)\to 0$ and $C'(r)\sim 2r$ while $\lambda(r)\sim s/(m r)$, so the spin corrections vanish at infinity and $\mathcal{E}$ and $\mathcal{J}$ coincide with the energy and azimuthal angular momentum measured at infinity.
Equations \eqref{3.18} and \eqref{3.19} form a linear algebraic system for $p_t$ and $p_\phi$. Solving it yields
\begin{equation}
p_{\phi}=\frac{\mathcal{J}-\frac{1}{2}C'(r)\lambda(r)\mathcal{E}}{\Delta(r)},\qquad
p_t=-\frac{\mathcal{E}-\frac{1}{2}A'(r)\lambda(r)\mathcal{J}}{\Delta(r)},
\label{3.20}
\end{equation}
where we have defined
\begin{equation}
\Delta(r)\equiv 1-\frac{1}{4}\,A'(r)\,C'(r)\,\lambda(r)^2.
\label{3.21}
\end{equation}
In the regime relevant for weak deflection of a pole-dipole test body, we will ultimately retain only terms linear in $s$. In that approximation, $\Delta(r)=1+\mathcal{O}(s^2)$ and Eq. \eqref{3.20} simplifies to
\begin{equation}
p_{\phi}=\mathcal{J}-\frac{1}{2}C'(r)\lambda(r)\mathcal{E}+\mathcal{O}(s^2),\qquad
p_t=-\mathcal{E}+\frac{1}{2}A'(r)\lambda(r)\mathcal{J}+\mathcal{O}(s^2).
\label{3.22}
\end{equation}
We now obtain the radial equation from the mass-shell constraint \eqref{3.6}, which reads on the equatorial plane
\begin{equation}
-\frac{p_t^2}{A(r)}+\frac{p_r^2}{B(r)}+\frac{p_\phi^2}{C(r)}=-m^2.
\label{3.23}
\end{equation}
Solving for $p_r^2$ and substituting Eq. \eqref{3.20} gives an exact radial first integral expressed solely in terms of $(\mathcal{E},\mathcal{J}\,m\,s)$ and the metric functions. For later analytic work it is also useful to display the linear-in-spin form, obtained by using Eq. \eqref{3.22} in Eq. \eqref{3.23}:
\begin{equation}
\frac{p_r^{\,2}}{B(r)}=
\frac{E^{2}}{A(r)}-\frac{J^{2}}{C(r)}-m^{2}
+\lambda(r)\,EJ\!\left(\frac{C'(r)}{C(r)}-\frac{A'(r)}{A(r)}\right)
+\mathcal{O}(s^{2}).
\label{3.24}
\end{equation}
To convert the radial first integral into an orbit equation $r(\phi)$, we must relate the tangent $u^\mu$ to the momentum $p^\mu$. Under the Tulczyjew-Dixon SSC, the momentum-velocity relation differs from proportionality only at quadratic order in spin, i.e. one has $p^\mu=m\,u^\mu+\mathcal{O}(s^2)$ for pole-dipole motion \cite{Semerak:1999qc,Costa:2017kdr}.
Therefore, to linear order in $s$ we may write
\begin{equation}
\frac{dr}{d\phi}=\frac{u^r}{u^\phi}=\frac{p^r}{p^\phi}+\mathcal{O}(s^2)
=\frac{g^{rr}p_r}{g^{\phi\phi}p_\phi}+\mathcal{O}(s^2)
=\frac{C(r)}{B(r)}\,\frac{p_r}{p_\phi}+\mathcal{O}(s^2).
\label{3.25}
\end{equation}
Squaring and using Eq. \eqref{3.23} to eliminate $p_r^2$ yields the orbit equation in the compact form
\begin{equation}
\left(\frac{dr}{d\phi}\right)^2=
\frac{C(r)^2}{B(r)\,p_\phi^2}\left(\frac{p_t^2}{A(r)}-m^2\right)-\frac{C(r)}{B(r)}
+\mathcal{O}(s^2),
\label{3.26}
\end{equation}
where $p_t(r)$ and $p_\phi(r)$ are given either exactly by Eq. \eqref{3.20} or, when working consistently to linear order in spin, by Eq. \eqref{3.22}.
Equation \eqref{3.26} is the orbit-level input we will use in the weak-deflection expansion: spin enters through the $r$-dependence of $p_t$ and $p_\phi$ induced by the spin terms in the conserved charges \eqref{3.18} and \eqref{3.19}. This dependence is ultimately responsible for the fact that the spatial trajectory is not a geodesic of the Jacobi geometry, even though it remains fully determined by the MPD system together with the SSC.

\section{Core theoretical result: Gauss-Bonnet with a non-geodesic particle ray} \label{sec4}
The geometric program reviewed in Section \ref{sec2} hinges on the fact that, for a spinless massive particle of fixed conserved energy, the spatial projection of the worldline is a geodesic of the two-dimensional Jacobi manifold. For a spinning particle governed by the MPD equations, this key property fails: spin-curvature coupling generically produces a non-geodesic spatial ray in the Jacobi geometry. In this section we formulate a Gauss-Bonnet construction in which the physical particle ray is allowed to be non-geodesic, while the auxiliary circular boundary is retained as a Jacobi geodesic, preserving the main simplification exploited by Li \textit{et al}.

\subsection{Lens domain \texorpdfstring{$D$}{} with \texorpdfstring{$\gamma_s$}{} non-geodesic and \texorpdfstring{$\gamma_{co}$}{} geodesic} \label{sec4.1}

We work on the equatorial Jacobi manifold $(\mathcal{M}J,\bar g_{ij})$ defined by the restriction of the Jacobi line element \eqref{2.4} to $\theta=\pi/2$, where $x^i=(r,\phi)$ and Latin indices $i,j,\ldots$ run over ${r,\phi}$. To avoid confusion with the four-dimensional spacetime metric $g_{\mu\nu}$, we denote the two-dimensional Jacobi metric by $\bar g_{ij}$ and its Levi-Civita connection by $\bar\nabla_i$. The arclength element induced by $\bar g_{ij}$ along a curve is denoted by $d\sigma$.
Let $\gamma_s$ be the spatial projection of the spinning particle trajectory in $\mathcal{M}J$, connecting a source point $S$ to a receiver point $R$. We parametrize $\gamma_s$ by Jacobi arclength $\sigma$ so that its unit tangent satisfies
\begin{equation}
T^i \equiv \frac{dx^i}{d\sigma},
\qquad
\bar g_{ij}T^iT^j=1.
\label{4.1}
\end{equation}
Because the MPD force is nonzero in general, $\gamma_s$ need not satisfy the Jacobi geodesic equation, and hence it will carry a nontrivial geodesic curvature. For any smooth, unit-speed curve $\gamma$ in $(\mathcal{M}J,\bar g_{ij})$ with unit normal $n^i$ chosen within the tangent space of $\mathcal{M}J$ and oriented toward the interior of the lens domain defined below, we define the geodesic curvature by
\begin{equation}
\kappa_g(\gamma)\equiv \bar g_{ij}\,(\bar\nabla_T T)^i\,n^j.
\label{4.2}
\end{equation}
With this convention, $\kappa_g(\gamma)=0$ if and only if $\gamma$ is a geodesic of $\bar g_{ij}$.
We now introduce the Gauss-Bonnet domain $D\subset\mathcal{M}J$. Fix coordinate locations for the source and receiver,
\begin{equation}
S:\ (r,\phi)=(r_S,\phi_S),
\qquad
R:\ (r,\phi)=(r_R,\phi_R),
\label{4.3}
\end{equation}
and define $\phi_{RS}\equiv \phi_R-\phi_S$ with $\phi_R>\phi_S$ under the standard orientation. We assume that the physical ray $\gamma_s$ remains in the region where the Jacobi metric is Riemannian and that it admits a single-valued representation $r=r(\phi)$ on the interval $\phi\in[\phi_S,\phi_R]$. We further assume the existence of a radius $r_{co}$ such that $r_{co}<r(\phi)$ throughout this interval; this ensures that the circular arc defined below lies entirely inside the physical trajectory in the Jacobi manifold and that the resulting domain is a topological disk.
The boundary $\partial D$ is taken to be the closed, piecewise smooth curve obtained by concatenating four segments: the physical ray $\gamma_s$ from $S$ to $R$; a radial segment $\gamma_R$ at fixed $\phi=\phi_R$ from $R$ down to the circle $r=r_{co}$; the circular arc $\gamma_{co}$ at fixed $r=r_{co}$ from $\phi=\phi_R$ back to $\phi=\phi_S$; and a radial segment $\gamma_S$ at fixed $\phi=\phi_S$ from the circle back up to $S$. The orientation of $\partial D$ is chosen to be positive (counterclockwise) so that the interior of $D$ lies to the left as one traverses $\partial D$.
The circular arc $\gamma_{co}$ is chosen to be a Jacobi geodesic, exactly as in Li \textit{et al}. Concretely, we impose
\begin{equation}
\kappa_g(\gamma_{co})=0,
\label{4.4}
\end{equation}
which is equivalent to the circular-orbit condition $\kappa(\gamma_{co})=0$ reviewed in Eq. \eqref{2.10} and its explicit form Eq. \eqref{2.11}. By contrast, the physical spinning-particle ray is not assumed to be geodesic \cite{Ono:2017pie,Werner:2012rc,Li:2019mqw}, and we allow
\begin{equation}
\kappa_g(\gamma_s)\neq 0
\label{4.5}
\end{equation}
in general.
At the vertices where the radial segments meet the circular arc, the boundary is orthogonal with respect to $\bar g_{ij}$. This follows because the Jacobi metric \eqref{2.4} is diagonal in $(r,\phi)$, so the tangents to $r=\mathrm{const}$ and $\phi=\mathrm{const}$ curves are orthogonal everywhere. Consequently, the two corner angles at these intersections contribute a total of $\pi$ to the sum of exterior jump angles in the Gauss-Bonnet theorem.
At the source and receiver vertices, we introduce the intrinsic angles $\Psi_S$ and $\Psi_R$ between the ray tangent and the outward radial direction. To define them, we introduce the outward-pointing unit radial vector in $\mathcal{M}J$,
\begin{equation}
e_r^i \equiv \frac{1}{\sqrt{\bar g_{rr}(r)}}\,(\partial_r)^i,
\label{4.6}
\end{equation}
evaluated at the relevant point. Let $T^i$ be the unit tangent to $\gamma_s$ at $S$ and $R$ oriented from $S$ to $R$. We then define $\Psi$ at either endpoint by
\begin{equation}
\cos\Psi \equiv \bar g_{ij}\,T^i e_r^j,
\qquad
0\le \Psi \le \pi,
\label{4.7}
\end{equation}
and set $\Psi_S\equiv \Psi|{S}$ and $\Psi_R\equiv \Psi|{R}$. These angles are purely geometric objects in the Jacobi manifold and reduce to the corresponding quantities used in the spinless construction when $\gamma_s$ becomes a Jacobi geodesic.
With the domain $D$ and the boundary data specified above, the Gauss-Bonnet theorem \eqref{2.13} can be applied without modification, but its boundary term will now receive a nontrivial contribution from the geodesic curvature integral along $\gamma_s$, reflecting the non-geodesic character of the spinning particle ray.

\subsection{Derivation of the spin-generalized deflection formula} \label{sec4.2}
We now derive the deflection-angle formula appropriate to a non-geodesic spinning-particle ray in the Jacobi manifold. The derivation follows the same Gauss-Bonnet logic as in the spinless case, with the crucial difference that the boundary contribution from the physical ray no longer vanishes.
We apply the Gauss-Bonnet theorem to the domain $D\subset\mathcal{M}J$ introduced in Section \ref{sec4.1}. Since $D$ is assumed to be a topological disk, its Euler characteristic is $\chi(D)=1$. Writing $K$ for the Gaussian curvature of $(\mathcal{M}J,\bar g_{ij})$, and $dS$ for the corresponding area element, the theorem gives \cite{Gibbons:2008rj,Werner:2012rc}
\begin{equation}
\iint_D K\,dS+\oint_{\partial D}\kappa_g\,d\sigma+\sum_{a=1}^{4}\beta_a=2\pi.
\label{4.8}
\end{equation}
Here $\kappa_g$ is the geodesic curvature defined in Eq. \eqref{4.2}, computed with the unit normal chosen toward the interior of $D$, and $\beta_a$ are the exterior jump angles at the four vertices of $\partial D$.
We next evaluate each term in Eq. \eqref{4.8} using the specific boundary decomposition
\begin{equation}
\partial D=\gamma_s\cup\gamma_R\cup\gamma_{co}\cup\gamma_S,
\label{4.9}
\end{equation}
with orientation as specified in Section \ref{sec4.1}. By construction, $\gamma_{co}$ is a Jacobi geodesic, so Eq. \eqref{4.4} implies $\kappa_g(\gamma_{co})=0$. The radial segments $\gamma_S$ and $\gamma_R$ are also Jacobi geodesics, because the Jacobi metric in Eq. \eqref{2.4} is diagonal and independent of $\phi$, implying that curves of constant $\phi$ satisfy the geodesic equation in $(\mathcal{M}J,\bar g_{ij})$. Therefore the boundary integral reduces to the contribution from the physical spinning-particle ray alone,
\begin{equation}
\oint_{\partial D}\kappa_g\,d\sigma=\int_{\gamma_s}\kappa_g\,d\sigma.
\label{4.10}
\end{equation}
We now compute the exterior jump-angle sum. At the two vertices where a radial segment meets the circular arc $\gamma_{co}$, the tangents are orthogonal with respect to $\bar g_{ij}$, so each vertex contributes an exterior angle of $\pi/2$, giving a total contribution $\pi$ from these two corners.
At the receiver $R$, the boundary transitions from $\gamma_s$ to the radial segment $\gamma_R$, whose tangent points inward along $-e_r^i$ at $R$. By Eq. \eqref{4.7}, the angle between the ray tangent and the outward radial direction $e_r^i$ is $\Psi_R$, and hence the angle between the ray tangent and the inward radial direction is $\pi-\Psi_R$. With the positive orientation of $\partial D$, this quantity is precisely the exterior jump angle at $R$, so we identify
\begin{equation}
\beta_R=\pi-\Psi_R.
\label{4.11}
\end{equation}
At the source $S$, the boundary transitions from the radial segment $\gamma_S$ to the ray $\gamma_s$. The tangent to $\gamma_S$ points outward along $+e_r^i$ at $S$, and by definition the angle between $+e_r^i$ and the ray tangent is $\Psi_S$. With the same orientation convention, this is the exterior jump angle at $S$, yielding
\begin{equation}
\beta_S=\Psi_S.
\label{4.12}
\end{equation}
Collecting all four vertices, we obtain
\begin{equation}
\sum_{a=1}^{4}\beta_a=\pi+\beta_S+\beta_R=\pi+\Psi_S+\pi-\Psi_R.
\label{4.13}
\end{equation}
Substituting Eqs. \eqref{4.10} and \eqref{4.13} into Eq. \eqref{4.8} and simplifying, we find the key relation between endpoint angles and curvature data:
\begin{equation}
\Psi_R-\Psi_S=\iint_D K\,dS+\int_{\gamma_s}\kappa_g\,d\sigma.
\label{4.14}
\end{equation}
This identity is purely geometric and holds for any smooth ray $\gamma_s$ connecting $S$ to $R$ in the Jacobi manifold, irrespective of whether $\gamma_s$ is a Jacobi geodesic.
Finally, we adopt the same finite-distance deflection-angle definition used in the spinless case, namely \cite{Ishihara:2016vdc,Takizawa:2020egm,Ono:2017pie}
\begin{equation}
\alpha \equiv \Psi_R-\Psi_S+\phi_{RS},
\qquad
\phi_{RS}\equiv \phi_R-\phi_S,
\label{4.15}
\end{equation}
where $\Psi_S$ and $\Psi_R$ are defined intrinsically by Eq. \eqref{4.7}. Combining Eq. \eqref{4.15} with Eq. \eqref{4.14} yields the spin-generalized Gauss-Bonnet deflection formula
\begin{equation}
\alpha=\iint_D K\,dS+\int_{\gamma_s}\kappa_g\,d\sigma+\phi_{RS}.
\label{4.16}
\end{equation}
In the spinless limit, the physical ray becomes a Jacobi geodesic and $\kappa_g(\gamma_s)\to 0$, so Eq. \eqref{4.16} reduces to the Li-type formula reviewed in Section \ref{sec2}. The new term $\int_{\gamma_s}\kappa_g\,d\sigma$ is therefore the unique additional contribution required to accommodate spin-induced non-geodesicity within the same Gauss-Bonnet lensing domain.

With the chosen counterclockwise orientation of $\partial D$, the interior-pointing normal on each positively oriented boundary segment coincides with the left normal used in Section \ref{sec5.1}, so $\kappa_g$ agrees with the signed curvature $k$ introduced there (restricted to $\bar g_{ij}$).

\subsection{Compatibility with Li's circular-orbit simplification} \label{sec4.3}
The spin-generalized deflection formula \eqref{4.16} contains two geometric contributions: the surface integral of the Gaussian curvature over the lens domain $D$, and the line integral of the geodesic curvature along the physical ray $\gamma_s$. The circular-orbit simplification of Li \textit{et al}. concerns only the surface term. Since that simplification relies exclusively on the intrinsic structure of the Jacobi manifold and on the choice of the lower boundary $\gamma_{co}$ as a Jacobi geodesic, it remains valid even when the upper boundary $\gamma_s$ is non-geodesic.
To make this explicit, we emphasize that the Jacobi manifold $(\mathcal{M}J,\bar g_{ij})$ is fixed once we choose the background SSS metric and the particle parameters entering the Jacobi conformal factor in Eq. \eqref{2.4}. In particular, the Gaussian curvature $K$ and the area element $dS$ are determined solely by $\bar g_{ij}$ and are insensitive to whether the physical ray is a Jacobi geodesic. For an SSS Jacobi metric, $\bar g_{ij}$ is diagonal in $(r,\phi)$ and depends only on $r$, so the curvature density $K\sqrt{\det\bar g}$ is a total radial derivative exactly as in Eq. \eqref{2.19} (with $g_{ij}\to \bar g_{ij}$). Accordingly, we use the same radial primitive introduced in Section \ref{sec2.4}, and denote it by $\bar{\mathcal{I}}(r)$ to stress that it is evaluated using $\bar g_{ij}$.
With the domain $D$ defined in Section \ref{sec4.1}, we may parameterize it by $\phi\in[\phi_S,\phi_R]$ and $r\in[r_{co},r(\phi)]$, where $r=r(\phi)$ describes the physical ray. Using the primitive $\bar{\mathcal{I}}(r)$ as in Eq. \eqref{2.20}, we obtain
\begin{equation}
\iint_{D}K\,dS
=\int_{\phi_S}^{\phi_R}\int_{r_{co}}^{r(\phi)} K\,\sqrt{\det \bar g}\,dr\,d\phi = \int_{\phi_S}^{\phi_R}\Big[\bar{\mathcal{I}}(r(\phi))-\bar{\mathcal{I}}(r_{co})\Big]\,d\phi.
\label{4.17}
\end{equation}
The role of the circular-orbit boundary is now identical to the spinless case. The identity in Eq. \eqref{2.22} implies that the primitive evaluated on a coordinate circle is proportional to the geodesic curvature of that circle (again with $g_{ij}\to\bar g_{ij}$). Since $\gamma_{co}$ is chosen to satisfy the Jacobi-geodesic condition \eqref{4.4}, it follows that $\bar{\mathcal{I}}(r_{co})=0$ after fixing the irrelevant additive constant in the primitive. Therefore Eq. \eqref{4.17} reduces to
\begin{equation}
\iint_{D}K\,dS=
\int_{\phi_S}^{\phi_R}\bar{\mathcal{I}}(r(\phi))\,d\phi.
\label{4.18}
\end{equation}
Substituting Eq. \eqref{4.18} into the spin-generalized Gauss-Bonnet formula \eqref{4.16} yields a compact representation in which Li’s surface-term simplification is preserved and the effect of spin enters only through the additional boundary contribution from the non-geodesic ray \cite{Ishihara:2016vdc,Crisnejo:2018uyn,Li:2019vhp}:
\begin{equation}
\alpha
=\int_{\phi_S}^{\phi_R}\bar{\mathcal{I}}(r(\phi))\,d\phi
+\int_{\gamma_s}\kappa_g\,d\sigma
+\phi_{RS}.
\label{4.19}
\end{equation}
For practical computations it is often convenient to express the line integral along $\gamma_s$ in terms of the coordinate parameter $\phi$. Writing $r=r(\phi)$ along the ray and using $d\sigma^2=\bar g_{rr}\,dr^2+\bar g_{\phi\phi}\,d\phi^2$, we have
\begin{equation}
\int_{\gamma_s}\kappa_g\,d\sigma
=\int_{\phi_S}^{\phi_R}
\kappa_g(\phi)\,
\sqrt{\bar g_{rr}(r(\phi))\left(\frac{dr}{d\phi}\right)^2+\bar g_{\phi\phi}(r(\phi))}\,d\phi.
\label{4.20}
\end{equation}
where $\kappa_g(\phi)$ denotes the geodesic curvature of the ray evaluated at the point labeled by $\phi$.
Equations \eqref{4.19} and \eqref{4.20} show the precise sense in which the Li \textit{et al}. circular-orbit device is compatible with a spinning, non-geodesic particle ray: the surface integral continues to collapse to a one-dimensional integral with no residual lower-bound term, while all deviations from the spinless construction are cleanly localized in the single additional boundary functional $\int_{\gamma_s}\kappa_g\,d\sigma$.

\section{Making the spin term computable: a working weak-field construction} \label{sec5}
The spin-generalized Gauss-Bonnet formula derived in Section \ref{sec4} isolates the effect of spin into a single additional boundary functional, the geodesic-curvature integral along the physical particle ray. In order to turn that term into a practical weak-field observable, we require an explicit expression for the geodesic curvature of an arbitrary curve in the two-dimensional Riemannian geometry relevant to the lensing construction. We provide that expression in this section in a form suited to perturbative evaluation along a ray given parametrically or as $r=r(\phi)$.

\subsection{Geodesic curvature \texorpdfstring{$k$}{} of a general curve in \texorpdfstring{$(\mathcal{M},g_{ij})$}{}} \label{sec5.1}
Let $(\mathcal{M},g_{ij})$ be an oriented two-dimensional Riemannian manifold with Levi-Civita connection $\nabla_i$. Consider a smooth curve $\gamma:\lambda\mapsto x^i(\lambda)$, with tangent
\begin{equation}
\dot x^i\equiv \frac{dx^i}{d\lambda},
\qquad
v\equiv \sqrt{g_{ij}\dot x^i\dot x^j},
\label{5.1}
\end{equation}
where $v$ is the speed with respect to $g_{ij}$. The arclength element is $d\sigma=v\,d\lambda$, and the unit tangent is $T^i=\dot x^i/v$.
On a two-dimensional oriented manifold there is a canonical antisymmetric Levi-Civita tensor $\varepsilon_{ij}$ satisfying $\nabla_k\varepsilon_{ij}=0$ and normalized in a positively oriented coordinate chart by $\varepsilon_{r\phi}=+\sqrt{\det g}$. Fixing this orientation, we define the signed geodesic curvature $k$ of $\gamma$ by requiring that the covariant acceleration of the unit tangent be purely normal,
\begin{equation}
(\nabla_T T)^i = k\,n^i,
\label{5.2}
\end{equation}
where $n^i$ is the unit normal obtained by a $+\pi/2$ rotation of $T^i$ according to the chosen orientation. With this convention, $k=0$ if and only if $\gamma$ is a geodesic of $g_{ij}$, and the sign of $k$ flips if the normal orientation is reversed.
A coordinate expression for $k$ that is directly suited to computation follows by writing the covariant acceleration in terms of the Christoffel symbols $\Gamma^{i}{}_{jk}$ of $g_{ij}$. Defining
\begin{equation}
A^i \equiv \frac{d^2 x^i}{d\lambda^2}+\Gamma^{i}{}_{jk}\frac{dx^j}{d\lambda}\frac{dx^k}{d\lambda},
\label{5.3}
\end{equation}
we obtain the compact formula
\begin{equation}
k=\frac{1}{v^3}\,\varepsilon_{ij}\,\dot x^i\,A^j.
\label{5.4}
\end{equation}
This identity is invariant under reparametrizations of the curve. Moreover, the combination that appears in the Gauss-Bonnet boundary term is $k\,d\sigma$, which can be written without explicit normalization as
\begin{equation}
k\,d\sigma=\frac{1}{v^2}\,\varepsilon_{ij}\,\dot x^i\,A^j\,d\lambda.
\label{5.5}
\end{equation}
For later use, we specialize Eq. \eqref{5.4} to the common situation in which $g_{ij}$ is diagonal and axisymmetric in coordinates $(r,\phi)$, namely
\begin{equation}
d\sigma^2=g_{rr}(r)\,dr^2+g_{\phi\phi}(r)\,d\phi^2,
\qquad
g_{rr}(r)>0,\quad g_{\phi\phi}(r)>0,
\label{5.6}
\end{equation}
and the curve is represented as $r=r(\phi)$ with $\phi$ used as a parameter. Writing primes for $d/dr$ and setting $r'\equiv dr/d\phi$, $r''\equiv d^2r/d\phi^2$, the speed becomes
\begin{equation}
v=\frac{d\sigma}{d\phi}=\sqrt{g_{rr}(r)\,r'^{\,2}+g_{\phi\phi}(r)}.
\label{5.7}
\end{equation}
A straightforward evaluation of Eq. \eqref{5.4} then yields the signed geodesic curvature
\begin{equation}
k=
\frac{\sqrt{g_{rr}(r)\,g_{\phi\phi}(r)}}{\big(g_{rr}(r)\,r'^{\,2}+g_{\phi\phi}(r)\big)^{3/2}}
\left[
-r''+
\left(\frac{g_{\phi\phi}'(r)}{g_{\phi\phi}(r)}-\frac{1}{2}\frac{g_{rr}'(r)}{g_{rr}(r)}\right)r'^{\,2}
+\frac{1}{2}\frac{g_{\phi\phi}'(r)}{g_{rr}(r)}
\right].
\label{5.8}
\end{equation}
Here $k$ denotes the \emph{signed} geodesic curvature with respect to the
chosen orientation of the $(r,\phi)$ coordinates and the Levi-Civita tensor
$\varepsilon_{ij}$ on $(\mathcal{M}J,\bar g_{ij})$, normalized by
$\varepsilon_{r\phi}=+\sqrt{\det \bar g}$.
Along an oriented curve $\gamma$ parametrized by $\lambda$, we take the unit
tangent $t^i\equiv \dot x^i/|\dot x|_{\bar g}$ and define the unit normal
$n^i\equiv \varepsilon^{i}{}_{j}\,t^{j}$ so that $(t,n)$ is positively oriented.
With this convention $k=\bar g_{ij}(\bar\nabla_t t)^i n^j$, and $k>0$ when the
curve turns toward the normal pointing to the \emph{left} of its direction of
motion.

As an internal consistency check, applying Eq. \eqref{5.8} to a coordinate circle $r=\mathrm{const}$ gives $r'=r''=0$ and hence
\begin{equation}
k\big|_{r=\mathrm{const}}=
\frac{1}{2\sqrt{g_{rr}(r)}}\,\frac{d}{dr}\ln g_{\phi\phi}(r)\,
\label{5.9}
\end{equation}
which is the standard geodesic-curvature formula for a coordinate circle in a diagonal axisymmetric metric and matches the expression used earlier when characterizing the circular-orbit boundary.

\subsection{Relating \texorpdfstring{$k$}{} to the spin-curvature force (MPD input)} \label{sec5.2}
We now connect the intrinsic geodesic curvature $k$ of the projected spinning-particle ray in the Jacobi manifold to the spin-curvature coupling that drives the motion away from a Jacobi geodesic. The outcome is a direct computational bridge: once the MPD force is known (to the desired perturbative order), the integrand $k\,d\sigma$ appearing in Eq. \eqref{4.19} can be evaluated along the ray using only local geometric data.
We consider the equatorial Jacobi manifold $(\mathcal{M}J,\bar g_{ij})$ with coordinates $(r,\phi)$ and line element
\begin{equation}
d\sigma^2=\bar g_{rr}(r)\,dr^2+\bar g_{\phi\phi}(r)\,d\phi^2,
\label{5.10}
\end{equation}
where $\bar g_{rr}$ and $\bar g_{\phi\phi}$ are the components of the Jacobi metric in Eq. \eqref{2.4} restricted to $\theta=\pi/2$. We parametrize the physical spinning-particle ray by proper time $\tau$ and write
\begin{equation}
x^i(\tau)=(r(\tau),\phi(\tau)),
\qquad
\dot x^i\equiv \frac{dx^i}{d\tau},
\qquad
v_J^2\equiv \bar g_{ij}\dot x^i\dot x^j,
\label{5.11}
\end{equation}
so that $d\sigma=v_J\,d\tau$. The geodesic curvature $k$ of the ray in the Jacobi manifold may then be computed from Eq. \eqref{5.5} with $\lambda=\tau$:
\begin{equation}
k\,d\sigma=\frac{1}{v_J^2}\,\bar\varepsilon_{ij}\,\dot x^i\,\bar A^j\,d\tau,
\qquad
\bar A^i\equiv \ddot x^i+\bar\Gamma^{i}{}_{jk}\dot x^j\dot x^k\,
\label{5.12}
\end{equation}
where $\bar\Gamma^{i}{}_{jk}$ are the Christoffel symbols of $\bar g_{ij}$ and $\bar\varepsilon_{ij}$ is the Levi-Civita tensor on $(\mathcal{M}J,\bar g_{ij})$, normalized by $\bar\varepsilon_{r\phi}=+\sqrt{\det\bar g}$ in the positively oriented chart.
The MPD input enters through the spacetime equation of motion. In the pole-dipole approximation with the Tulczyjew-Dixon SSC, we define the MPD force by
\begin{equation}
f^\mu\equiv \frac{D p^\mu}{d\tau}=-\frac{1}{2}\,R^\mu{}_{\nu\alpha\beta}\,u^\nu S^{\alpha\beta},
\label{5.13}
\end{equation}
and the four-acceleration by $a^\mu\equiv D u^\mu/d\tau$. In the regime in which we retain only terms linear in the spin, the momentum-velocity relation implies $p^\mu=m u^\mu+\mathcal{O}(s^2)$, hence
\begin{equation}
a^\mu=\frac{1}{m}\,f^\mu +\mathcal{O}(s^2)=
-\frac{1}{2m}\,R^\mu{}_{\nu\alpha\beta}\,u^\nu S^{\alpha\beta}+\mathcal{O}(s^2).
\label{5.14}
\end{equation}
This is the precise sense in which the deviation from geodesic motion is driven by spin-curvature coupling at leading order \cite{Semerak:1999qc,Costa:2017kdr,Dixon:1974xoz,Compere:2021kjz}.

To use Eq. \eqref{5.14} in Eq. \eqref{5.12}, we relate the coordinate second derivatives $\ddot r$ and $\ddot\phi$ to the spacetime four-acceleration components $a^r$ and $a^\phi$. For the SSS spacetime metric \eqref{2.1}, restricted to the equatorial plane and using proper time as parameter, we have
\begin{equation}
a^r=\ddot r+\Gamma^{r}{}_{tt}\,(u^t)^2+\Gamma^{r}{}_{rr}\,(u^r)^2+\Gamma^{r}{}_{\phi\phi}\,(u^\phi)^2,
\label{5.15}
\end{equation}
\begin{equation}
a^\phi=\ddot\phi+2\,\Gamma^{\phi}{}_{r\phi}\,u^r u^\phi\,
\label{5.16}
\end{equation}
where $u^r=\dot r$ and $u^\phi=\dot\phi$. For the diagonal SSS line element \eqref{2.1}, the required Christoffel symbols on $\theta=\pi/2$ are
\begin{equation}
\Gamma^{r}{}_{tt}=\frac{A'(r)}{2B(r)},
\qquad
\Gamma^{r}{}_{rr}=\frac{B'(r)}{2B(r)},
\qquad
\Gamma^{r}{}_{\phi\phi}=-\frac{C'(r)}{2B(r)},
\qquad
\Gamma^{\phi}{}_{r\phi}=\frac{C'(r)}{2C(r)}.
\label{5.17}
\end{equation}
Solving Eqs. \eqref{5.15} and \eqref{5.16} for $\ddot r$ and $\ddot\phi$ and substituting into Eq. \eqref{5.12} yields $\bar A^i$ expressed in terms of the spacetime acceleration $(a^r\,a^\phi)$ and known connection coefficients.
For the Jacobi metric itself, Eq. \eqref{5.10} implies that the only nonvanishing $\bar\Gamma^{i}{}_{jk}$ are
\begin{equation}
\bar\Gamma^{r}{}_{rr}=\frac{1}{2}\frac{d}{dr}\ln \bar g_{rr}(r)\,
\qquad
\bar\Gamma^{r}{}_{\phi\phi}=-\frac{1}{2\bar g_{rr}(r)}\frac{d\bar g_{\phi\phi}(r)}{dr},
\qquad
\bar\Gamma^{\phi}{}_{r\phi}=\frac{1}{2}\frac{d}{dr}\ln \bar g_{\phi\phi}(r).
\label{5.18}
\end{equation}
Using Eqs. \eqref{5.15}-\eqref{5.18}, we obtain the Jacobi covariant acceleration components along the spinning ray in the form
\begin{equation}
\bar A^{r}=
a^{r}
-\Gamma^{r}{}_{tt}\,(u^t)^2
+\big(\bar\Gamma^{r}{}_{rr}-\Gamma^{r}{}_{rr}\big)\,(u^{r})^{2}
+\big(\bar\Gamma^{r}{}_{\phi\phi}-\Gamma^{r}{}_{\phi\phi}\big)\,(u^{\phi})^{2},
\label{5.19}
\end{equation}
\begin{equation}
\bar A^{\phi}=
a^{\phi}
+2\big(\bar\Gamma^{\phi}{}_{r\phi}-\Gamma^{\phi}{}_{r\phi}\big)\,u^{r}u^{\phi}.
\label{5.20}
\end{equation}
These identities are exact kinematic relations for any worldline parametrized by $\tau$; all dynamical content enters through $a^\mu$, which is supplied by the MPD force via Eq. \eqref{5.14} at linear order in spin.
Finally, it is useful to write the integrand $k\,d\sigma$ appearing in Eq. \eqref{4.19} in a compact coordinate form. Since $\bar\varepsilon_{r\phi}=+\sqrt{\det\bar g}$, Eq. \eqref{5.12} gives
\begin{equation}
k\,d\sigma=
\frac{\sqrt{\det\bar g}}{v_J^{2}}\Big(\dot r\,\bar A^{\phi}-\dot\phi\,\bar A^{r}\Big)\,d\tau,
\qquad
v_J^{2}=\bar g_{rr}\dot r^{\,2}+\bar g_{\phi\phi}\dot\phi^{\,2}.
\label{5.21}
\end{equation}
Equations \eqref{5.14}, \eqref{5.19}, \eqref{5.20}, and \eqref{5.21} constitute the promised bridge between the MPD description and the Gauss-Bonnet spin term. In practice, one computes $u^t$, $u^r$, and $u^\phi$ from the effective first integrals in Section \ref{sec3} (to the desired weak-field and small-spin order), evaluates the curvature contraction $R^\mu{}_{\nu\alpha\beta}u^\nu S^{\alpha\beta}$ to obtain $a^r$ and $a^\phi$ through Eq. \eqref{5.14}, and then substitutes into Eq. \eqref{5.21} to obtain $k\,d\sigma$ along the ray.

\subsection{Perturbative expansion strategy (what gets expanded, and what does not)} \label{sec5.3}
Our starting point for computation is the spin-generalized Gauss-Bonnet representation
\begin{equation}
\alpha=
\int_{\phi_S}^{\phi_R}\bar{\mathcal{I}}(r(\phi))\,d\phi
+
\int_{\gamma_s}k\,d\sigma
+
\phi_{RS},
\label{5.22}
\end{equation}
which is Eq. \eqref{4.19} written with the notation of Section \ref{sec5}. Here $\bar{\mathcal{I}}$ is the Li-type radial primitive built from the Jacobi metric, $r(\phi)$ is the physical spinning-particle ray, and the second term is the geodesic-curvature functional whose MPD reduction was given in Eq. \eqref{5.21}. The weak-field recipe is a controlled expansion of each term in Eq. \eqref{5.22} in small, dimensionless parameters characterizing the background field strength and the intrinsic spin.
We assume that the lensing configuration lies in a region where the SSS metric functions in Eq. \eqref{2.1} are close to their flat-space values in the sense that the dimensionless potentials evaluated at the scale of the impact parameter $b$ are small. For the spacetimes of interest later (Schwarzschild, Reissner-Nordstr\"om, and Kottler), this corresponds to the simultaneous smallness of
\begin{equation}
\epsilon_M\equiv \frac{M}{b}\ll 1,\qquad
\epsilon_Q\equiv \frac{Q^2}{b^2}\ll 1,\qquad
\epsilon_\Lambda\equiv |\Lambda|\,b^2\ll 1,
\label{5.23}
\end{equation}
together with the analogous conditions at the finite source and receiver radii, $M/r_{S,R}\ll 1$, $Q^2/r_{S,R}^2\ll 1$, and $|\Lambda|\,r_{S,R}^2\ll 1$ when $\Lambda\neq 0$. Independently, we work at linear order in the spin,
\begin{equation}
\epsilon_s\equiv \frac{s}{m\,b}\ll 1,
\label{5.24}
\end{equation}
and we consistently neglect $\mathcal{O}(s^2)$ effects, in particular the momentum-velocity misalignment beyond Eq. \eqref{5.14}. In practice we will retain terms through a specified combined order in the set ${\epsilon_M,\epsilon_Q,\epsilon_\Lambda,\epsilon_s}$; the key point is that the spin contribution is organized as an expansion in $\epsilon_s$ with coefficients that are themselves weak-field series.
A crucial conceptual point is what we do \emph{not} expand. We keep the Gauss-Bonnet identity itself exact, including the finite-distance definition of $\alpha$ and the topological characterization of the domain. We also keep the conserved particle parameters $m$ and the Killing charges $(\mathcal{E},\mathcal{J})$ fixed, treating the asymptotic speed (equivalently $\mathcal{E}/m$) and the impact parameter $b$ as external control parameters. In particular, the Jacobi metric is taken to be the same energy-dependent metric as in Section \ref{sec2}, with the energy parameter identified with the conserved energy at infinity of the spinning particle; this identification is natural because the Killing charge $\mathcal{E}$ defined in Eq. \eqref{3.10} coincides with $-p_t$ asymptotically, where the spin correction vanishes, and therefore labels the physical beam in the same way as in the spinless case.
By contrast, what we \emph{do} expand are the integrands in Eq. \eqref{5.22} and, when higher accuracy is desired, the ray shape $r(\phi)$ that feeds those integrands. The organizing principle is a Born-type hierarchy based on power counting in the weak-field parameters. The primitive $\bar{\mathcal{I}}(r)$ is constructed from the Jacobi metric and therefore admits a weak-field expansion of the form
\begin{equation}
\bar{\mathcal{I}}(r)=\bar{\mathcal{I}}^{(1)}(r)+\bar{\mathcal{I}}^{(2)}(r)+\cdots,
\qquad
\bar{\mathcal{I}}^{(n)}(r)=\mathcal{O}(\epsilon^n),
\label{5.25}
\end{equation}
where $\epsilon$ stands for any of $\epsilon_M,\epsilon_Q,\epsilon_\Lambda$ with the appropriate dimensions absorbed by powers of $r$ or $b$. Similarly, the geodesic-curvature integrand $k\,d\sigma$ is linear in the Jacobi covariant acceleration of the ray, and hence linear in the MPD force at leading order. Using Eq. \eqref{5.14} in Eq. \eqref{5.21} implies the scaling
\begin{equation}
k\,d\sigma = \mathcal{O}(\epsilon_s,\epsilon)\,d(\text{length}),
\label{5.26}
\end{equation}
so that the spin correction to the deflection angle begins at combined order $\mathcal{O}(\epsilon_s,\epsilon)$.
For the ray itself, we expand around the straight-line trajectory in the auxiliary flat-space polar geometry. Introducing $u(\phi)\equiv 1/r(\phi)$, the unperturbed scattering ray is
\begin{equation}
u_0(\phi)=\frac{\sin\phi}{b},
\label{5.27}
\end{equation}
with endpoint angles fixed by the Euclidean relations
\begin{equation}
\phi_S^{(0)}=\arcsin\!\left(\frac{b}{r_S}\right),\qquad
\phi_R^{(0)}=\pi-\arcsin\!\left(\frac{b}{r_R}\right).
\label{5.28}
\end{equation}
We then seek $u(\phi)$ as a perturbation series driven by the orbit equation \eqref{3.26} together with the linear-in-spin first integrals \eqref{3.22}, of the form
\begin{equation}
u(\phi)=u_0(\phi)+u_{\text{wf}}(\phi)+u_{\text{spin}}(\phi)+\cdots,
\qquad
u_{\text{wf}}=\mathcal{O}(\epsilon),\quad
u_{\text{spin}}=\mathcal{O}(\epsilon_s,\epsilon),
\label{5.29}
\end{equation}
with higher-order corrections included as needed.
This expansion induces a corresponding hierarchy for Eq. \eqref{5.22}. At leading weak-field order, we evaluate the surface term by substituting the unperturbed ray $r_0(\phi)=1/u_0(\phi)$ into $\bar{\mathcal{I}}^{(1)}(r)$ and integrating from $\phi_S^{(0)}$ to $\phi_R^{(0)}$. Corrections from replacing $r_0$ by $r=r_0+\delta r$ produce contributions suppressed by at least one extra weak-field order because they enter through a Taylor expansion,
\begin{equation}
\bar{\mathcal{I}}(r(\phi))=\bar{\mathcal{I}}(r_0(\phi))+\left.\frac{d\bar{\mathcal{I}}}{dr}\right|_{r_0(\phi)}\delta r(\phi)+\cdots,
\label{5.30}
\end{equation}
and $\delta r=\mathcal{O}(\epsilon)$ or $\mathcal{O}(\epsilon_s,\epsilon)$. Likewise, shifts of the integration limits away from Eq. \eqref{5.28} contribute at higher order because they multiply integrands that are already $\mathcal{O}(\epsilon)$ or $\mathcal{O}(\epsilon_s,\epsilon)$. This is the standard rationale for evaluating curvature integrals on an unperturbed domain at the lowest nontrivial order.
For the spin term, the same power counting is even more favorable. Since $k\,d\sigma$ is already $\mathcal{O}(\epsilon_s,\epsilon)$ by Eq. \eqref{5.26}, it is consistent at that order to compute it using the zeroth-order ray kinematics $(r_0(\phi),\dot r_0,\dot\phi_0)$ inside Eq. \eqref{5.21}, while evaluating the MPD force in Eq. \eqref{5.14} on the corresponding spinless background motion. Corrections from using a weak-field-corrected ray inside $k\,d\sigma$ would contribute at $\mathcal{O}(\epsilon_s,\epsilon^2)$ or beyond and can be included only if that combined accuracy is required.
In summary, our perturbative strategy is to treat Eq. \eqref{5.22} as an expansion problem in which the Jacobi-geometry quantities $(\bar{\mathcal{I}},K)$ are expanded in weak-field parameters, the non-geodesic functional $k\,d\sigma$ is evaluated using the MPD force at linear order in spin, and the ray $r(\phi)$ is improved iteratively only to the order demanded by the targeted accuracy of $\alpha$.

\subsection{Finite-distance angles with spin (keeping Li's structure)} \label{sec5.4}
The finite-distance deflection angle is defined geometrically by Eq. \eqref{4.15},
\begin{equation}
\alpha\equiv \Psi_R-\Psi_S+\phi_{RS},
\label{5.31}
\end{equation}
with $\Psi_S$ and $\Psi_R$ defined intrinsically in the Jacobi manifold by Eq. \eqref{4.7}. This definition does not rely on the ray being a Jacobi geodesic, and we therefore keep it unchanged in the spinning case. The only spin dependence enters through the direction of the physical ray at the endpoints, i.e. through the tangent $T^i$ of $\gamma_s$ at $S$ and $R$.
To obtain a computable expression, we consider the Jacobi metric on the equatorial plane in diagonal form,
\begin{equation}
d\sigma^2=\bar g_{rr}(r)\,dr^2+\bar g_{\phi\phi}(r)\,d\phi^2,
\label{5.32}
\end{equation}
and represent the ray locally as $r=r(\phi)$. Using the definition \eqref{4.7} with the outward unit radial vector $e_r^i=(\partial_r)^i/\sqrt{\bar g_{rr}}$ and the unit tangent $T^i$, a direct computation yields
\begin{equation}
\sin\Psi=
\frac{\sqrt{\bar g_{\phi\phi}(r)}}{\sqrt{\bar g_{rr}(r)\left(\frac{dr}{d\phi}\right)^2+\bar g_{\phi\phi}(r)}},
\qquad
\cos\Psi=
\frac{\sqrt{\bar g_{rr}(r)}\left(\frac{dr}{d\phi}\right)}{\sqrt{\bar g_{rr}(r)\left(\frac{dr}{d\phi}\right)^2+\bar g_{\phi\phi}(r)}},
\label{5.33}
\end{equation}
with $\Psi=\Psi_S$ or $\Psi_R$ when evaluated at $r=r_S$ or $r=r_R$, respectively. For the Jacobi metric \eqref{2.4} one has $\bar g_{rr}=F(r)B(r)$ and $\bar g_{\phi\phi}=F(r)C(r)$ with the same positive conformal factor $F(r)\equiv (E^2-m^2A)/A$, so the conformal factor cancels in the angle, giving the particularly simple relation
\begin{equation}
\tan\Psi=
\sqrt{\frac{C(r)}{B(r)}}\left|\frac{d\phi}{dr}\right|.
\label{5.34}
\end{equation}
The absolute value $|d\phi/dr|$ appears because $\Psi$ is defined as the
\emph{unsigned} angle between the ray tangent and the outward radial direction
in the Jacobi (optical) geometry at the endpoints. Equivalently, $\Psi\in(0,\pi)$
is determined by the magnitudes of the projected tangent components, while the
overall orientation information is carried separately by $\phi_{RS}\equiv
\phi_R-\phi_S$ and by the chosen (counterclockwise) orientation of $\partial D$
in the Gauss-Bonnet construction. In particular, changing the direction of
radial monotonicity near an endpoint flips the sign of $d\phi/dr$ but does not
change $\Psi$. Thus the finite-distance angle is controlled only by the coordinate slope of the ray and by the background functions $B(r)$ and $C(r)$.
It is sometimes useful to eliminate the slope in favor of momentum data. Using $dr/d\phi=u^r/u^\phi$ and the linear-in-spin identification $u^\mu=p^\mu/m+\mathcal{O}(s^2)$ together with the orbit relation \eqref{3.26}, Eq. \eqref{5.33} can be rewritten as
\begin{equation}
\sin^2\Psi=
\frac{p_\phi^{\,2}}{C(r)\left(\frac{p_t^{\,2}}{A(r)}-m^2\right)}+\mathcal{O}(s^2),
\label{5.35}
\end{equation}
where $p_t(r)$ and $p_\phi(r)$ are understood as the momentum components along the trajectory at the point of interest. Substituting the linear-in-spin expressions \eqref{3.22} into Eq. \eqref{5.35} and expanding consistently to first order in $s$ yields
\begin{equation}
\sin\Psi=
\frac{\mathcal{J}}{\sqrt{C(r)}\sqrt{\frac{\mathcal{E}^2}{A(r)}-m^2}}
+
\frac{\lambda(r)\,\mathcal{E}}{2\sqrt{C(r)}\sqrt{\frac{\mathcal{E}^2}{A(r)}-m^2}}
\left[
-C'(r)
+\frac{A'(r)\,\mathcal{J}^2}{A(r)\left(\frac{\mathcal{E}^2}{A(r)}-m^2\right)}
\right]
+\mathcal{O}(s^2),
\label{5.36}
\end{equation}
where $\mathcal{E}$ and $\mathcal{J}$ are the Killing charges defined in Eq. \eqref{3.10} and $\lambda(r)$ is the aligned-spin coupling introduced in Eq. \eqref{3.16}. Equation \eqref{5.26} provides an explicit finite-distance spin correction to the endpoint angles in terms of local metric data and conserved charges, while preserving the same geometric definition of $\Psi$ used in the spinless case.
Finally, for asymptotically flat spacetimes the spin-dependent pieces vanish at infinity because $A'(r)\to 0$ and $\lambda(r)\to 0$ as $r\to\infty$. In that regime we may define the impact parameter $b$ by the same asymptotic matching as in the spinless case,
\begin{equation}
\mathcal{J}=b\,\sqrt{\mathcal{E}^2-m^2},
\label{5.37}
\end{equation}
and Eq. \eqref{5.35} reduces to $\sin\Psi\simeq b/r$ at large $r$, recovering the expected flat-space behavior of the finite-distance angle.

\subsection{Final implementation-ready master formula (generic SSS)} \label{sec5.5}
We now collect the results of Sections 3-5 into a single master formula for the finite-distance deflection angle of an \emph{aligned-spin} massive test particle in a generic static spherically symmetric spacetime. Throughout, we work consistently to linear order in the spin (neglecting $\mathcal{O}(s^2)$), as assumed in Section \ref{sec5.3}.
We start from the generic SSS spacetime metric
\begin{equation}
ds^2=-A(r)\,dt^2+B(r)\,dr^2+C(r)\Big(d\theta^2+\sin^2\theta\,d\phi^2\Big),
\label{5.38}
\end{equation}
and restrict to equatorial motion $\theta=\pi/2$. We identify the energy parameter appearing in the Jacobi conformal factor with the conserved Killing energy of the spinning body, i.e. we set $E\equiv \mathcal{E}$. The (equatorial) Jacobi metric then takes the diagonal form
\begin{equation}
d\sigma^2=\bar g_{rr}(r)\,dr^2+\bar g_{\phi\phi}(r)\,d\phi^2,
\qquad
\bar g_{rr}(r)=F(r)\,B(r)\,\quad \bar g_{\phi\phi}(r)=F(r)\,C(r)\,
\label{5.39}
\end{equation}
with the conformal factor
\begin{equation}
F(r)\equiv \frac{E^2-m^2A(r)}{A(r)}.
\label{5.40}
\end{equation}
The circular boundary radius $r_{co}$ is fixed by the Jacobi-geodesic condition (equivalently $\kappa_g(\gamma_{co})=0$) stated in Eq. \eqref{2.11}, and all quantities are assumed to be evaluated in the region where $F(r)>0$.
The deflection angle is computed from the spin-generalized Gauss-Bonnet formula \eqref{4.19}, which we now write in a directly usable form:
\begin{equation}
\alpha=
\phi_{RS}
+\int_{\phi_S}^{\phi_R}\bar{\mathcal{I}}(r(\phi))\,d\phi
+\int_{\gamma_s} k\,d\sigma,
\qquad
\phi_{RS}\equiv \phi_R-\phi_S.
\label{5.41}
\end{equation}
Here $r=r(\phi)$ is the physical spinning-particle ray in the Jacobi manifold, $\bar{\mathcal{I}}(r)$ is the Li-type radial primitive built from the Jacobi metric, and the final term is the geodesic-curvature contribution of the non-geodesic ray.
The circular-orbit simplification fixes the surface contribution entirely in terms of $\bar{\mathcal{I}}(r)$ evaluated on the physical ray. Using Eq. \eqref{2.21} with $g_{ij}\to \bar g_{ij}$, one convenient closed form is
\begin{equation}
\bar{\mathcal{I}}(r)=
-\frac{1}{2}\sqrt{\frac{C(r)}{B(r)}}\,
\frac{d}{dr}\ln\!\big(F(r)\,C(r)\big),
\label{5.42}
\end{equation}
or, upon inserting Eq. \eqref{5.40},
\begin{equation}
\bar{\mathcal{I}}(r)=
-\frac{1}{2}\sqrt{\frac{C(r)}{B(r)}}\left[
\frac{C'(r)}{C(r)}
-\frac{E^2\,A'(r)}{A(r)\big(E^2-m^2A(r)\big)}
\right].
\label{5.43}
\end{equation}
These expressions are valid for any SSS metric in the form \eqref{5.38} and require no assumption about the ray being geodesic.
The spin contribution is encoded in the line integral of the geodesic curvature. An explicitly geometric representation, suited to a ray written as $r=r(\phi)$, is
\begin{equation}
\int_{\gamma_s}k\,d\sigma=
\int_{\phi_S}^{\phi_R}
k(\phi)\,
\sqrt{\bar g_{rr}(r(\phi))\left(\frac{dr}{d\phi}\right)^2+\bar g_{\phi\phi}(r(\phi))}\,d\phi\,
\label{5.44}
\end{equation}
where $k(\phi)$ can be computed from Eq. \eqref{5.8} by setting $g_{ij}=\bar g_{ij}$.
For practical implementation, it is often preferable to compute $k\,d\sigma$ directly from the MPD force without differentiating the orbit twice. Retaining only $\mathcal{O}(s)$ terms, we use $p^\mu=m u^\mu+\mathcal{O}(s^2)$ and the MPD acceleration \eqref{5.14} together with the bridge formula \eqref{5.21} to write
\begin{equation}
\int_{\gamma_s}k\,d\sigma=
\int_{\tau_S}^{\tau_R}
\frac{\sqrt{\det\bar g}}{v_J^{2}}
\Big(\dot r\,\bar A^{\phi}-\dot\phi\,\bar A^{r}\Big)\,d\tau,
\qquad
v_J^{2}=\bar g_{rr}\dot r^{\,2}+\bar g_{\phi\phi}\dot\phi^{\,2},
\label{5.45}
\end{equation}
with $\sqrt{\det\bar g}=\sqrt{\bar g_{rr}\bar g_{\phi\phi}}$. The Jacobi covariant-acceleration components $\bar A^{r}$ and $\bar A^{\phi}$ are related to the spacetime acceleration $(a^{r}\,a^{\phi})$ by Eqs. \eqref{5.19} and \eqref{5.20}, while the spacetime acceleration is fixed by the MPD force,
\begin{equation}
a^\mu=
-\frac{1}{2m}\,R^\mu{}_{\nu\alpha\beta}\,u^\nu S^{\alpha\beta}
+\mathcal{O}(s^2),
\label{5.46}
\end{equation}
evaluated on the aligned-spin sector specified in Section \ref{sec3.2}.
To close the system, the ray kinematics $(r(\phi),u^\mu,S^{\mu\nu})$ entering Eqs. \eqref{5.41} and \eqref{5.45} are determined by the effective first integrals of Section \ref{sec3}. In the aligned-spin equatorial sector, we introduce
\begin{equation}
\lambda(r)\equiv \frac{\sigma\,s}{m\sqrt{A(r)\,B(r)\,C(r)}},
\qquad
\sigma=\pm 1,
\label{5.47}
\end{equation}
and use the linear-in-spin relations \eqref{3.22},
\begin{equation}
p_\phi(r)=\mathcal{J}-\frac{1}{2}\,C'(r)\,\lambda(r)\,E+\mathcal{O}(s^2),
\qquad
p_t(r)=-E+\frac{1}{2}\,A'(r)\,\lambda(r)\,\mathcal{J}+\mathcal{O}(s^2),
\label{5.48}
\end{equation}
together with the mass-shell constraint \eqref{3.23} to obtain $p_r(r)$ and hence the orbit equation \eqref{3.26} for $r(\phi)$ at the same order. When evaluating the MPD force in Eq. \eqref{5.46}, the spin tensor can be expressed in terms of $p_\mu$ and the aligned spin four-vector through Eq. \eqref{3.13}, with the alignment encoded by Eq. \eqref{3.14}.
Equations \eqref{5.41}-\eqref{5.48} comprise an implementation-ready master system for generic SSS spacetimes: the surface term is obtained from the primitive \eqref{5.42} or \eqref{5.43} evaluated along the ray, and the spin correction is obtained from either the purely geometric representation \eqref{5.44} or the MPD-driven representation \eqref{5.45}-\eqref{5.46}, with the ray and spin variables determined by the first integrals \eqref{5.48} and the orbit equation \eqref{3.26} to linear order in $s$.

\section{Application} \label{sec6}
In this section we apply the master formula assembled in Section \ref{sec5.5} to concrete SSS metrics. We begin with Schwarzschild as a validation case, focusing on two checks: (i) the spinless limit reproduces the standard weak deflection with finite-distance dependence in Li’s framework, and (ii) the spin-dependent contribution enters with the expected weak-field scaling dictated by the MPD spin-curvature force.

\subsection{Schwarzschild case} \label{sec6.1}
We take the Schwarzschild spacetime of mass $M$,
\begin{equation}
A(r)=1-\frac{2M}{r},\qquad B(r)=\frac{1}{A(r)},\qquad C(r)=r^2.
\label{6.1}
\end{equation}
We parameterize the conserved Killing energy by $E \equiv \mathcal{E}=m\gamma$ and define the asymptotic speed $v$ by $\gamma=(1-v^2)^{-1/2}$. The asymptotic spatial momentum magnitude is
\begin{equation}
p_\infty\equiv \sqrt{E^2-m^2}=m\gamma v,
\label{6.2}
\end{equation}
and we define the impact parameter by
\begin{equation}
b\equiv \frac{\mathcal{J}}{p_\infty}.
\label{6.3}
\end{equation}
In the spinless limit, the ray is a Jacobi geodesic and the non-geodesic boundary term in Eq. \eqref{5.41} vanishes. At leading order in $M/b$ (and at finite $r_S,r_R$), Eq. \eqref{5.41} reduces to the standard weak deflection for a massive particle, which in Li’s/Ishihara’s finite-distance formulation can be written as
\begin{equation}
\alpha_{\rm Schw}^{(0)}=
\left(1+\frac{1}{v^2}\right)\frac{M}{b}
\left(
\sqrt{1-\frac{b^2}{r_S^2}}
+
\sqrt{1-\frac{b^2}{r_R^2}}
\right)
+\mathcal{O}\!\left(\frac{M^2}{b^2}\right),
\label{6.4}
\end{equation}
with $\sin\phi_S=b/r_S$ and $\sin(\pi-\phi_R)=b/r_R$ defining the endpoint angles in the Born approximation. Since this term is well documented, we now focus on the spin contribution, for which our Gauss-Bonnet synthesis supplies a distinct, computable boundary functional.
We work in the aligned-spin equatorial sector of Section \ref{sec3} and keep only terms linear in the spin length $s$, consistently neglecting $\mathcal{O}(s^2)$. We encode the spin orientation by $\sigma=\pm 1$, with $\sigma=+1$ corresponding to spin aligned with the orbital angular momentum and $\sigma=-1$ to anti-alignment. The full deflection angle is then
\begin{equation}
\alpha_{\rm Schw}=\alpha_{\rm Schw}^{(0)}+\alpha_{\rm Schw}^{(s)}+\mathcal{O}\!\left(\frac{M^2}{b^2}\right){\rm spinless} + \mathcal{O}(s^2),
\label{6.5}
\end{equation}
where the spin correction is precisely the non-geodesic boundary term from Eq. \eqref{5.41},
\begin{equation}
\alpha_{\rm Schw}^{(s)}=\int_{\gamma_s}k\,d\sigma.
\label{6.6}
\end{equation}
We now evaluate Eq. \eqref{6.6} transparently using the MPD-to-$k$ bridge developed in Section \ref{sec5}.2. We start from Eq. \eqref{5.45},
\begin{equation}
\int_{\gamma_s}k\,d\sigma=
\int_{\tau_S}^{\tau_R}
\frac{\sqrt{\det\bar g}}{v_J^{2}}
\Big(u^r\,\bar A^{\phi}-u^\phi\,\bar A^{r}\Big)\,d\tau,
\qquad
v_J^{2}=\bar g_{rr}(u^r)^2+\bar g_{\phi\phi}(u^\phi)^2,
\label{6.7}
\end{equation}
where $\bar g_{ij}$ is the Jacobi metric and $u^i=dx^i/d\tau$ are the spatial components of the four-velocity. We emphasize that Eq. \eqref{6.7} is exact and parameter-invariant.
We next apply the perturbative strategy of Section \ref{sec5}.3. Since $\alpha_{\rm Schw}^{(s)}$ is already $\mathcal{O}(s)$, it suffices at leading weak-field order to evaluate Eq. \eqref{6.7} on the unperturbed (spinless) scattering kinematics. In particular, for $s=0$ the ray is a Jacobi geodesic and hence $k=0$; equivalently, the combination $u^r\bar A^\phi-u^\phi\bar A^r$ in Eq. \eqref{6.7} vanishes identically at $s=0$. Therefore, to linear order in $s$ we may replace $\bar A^i$ in Eq. \eqref{6.7} by its $\mathcal{O}(s)$ part. At leading weak-field order, that $\mathcal{O}(s)$ part is sourced solely by the MPD acceleration, so we may use
\begin{equation}
u^r\bar A^\phi-u^\phi\bar A^r=
u^r a^\phi-u^\phi a^r
+\mathcal{O}\!\left(s\,\frac{M^2}{r^5}\right),
\label{6.8}
\end{equation}
where $a^\mu=D u^\mu/d\tau$ is fixed by the MPD force (Eq. \eqref{5.46}).
We now simplify the prefactor $\sqrt{\det\bar g}/v_J^2$ for Schwarzschild. From Eqs. \eqref{5.39}-\eqref{5.40} and Eq. \eqref{6.1}, the equatorial Jacobi metric reads $\bar g_{rr}=F(r)B(r)$ and $\bar g_{\phi\phi}=F(r)r^2$ with $F(r)=(E^2-m^2A)/A$. Hence
\begin{equation}
\frac{\sqrt{\det\bar g}}{v_J^2}=\frac{\sqrt{\bar g_{rr}\bar g_{\varphi\varphi}}}
{\bar g_{rr}(u^{r})^{2}+\bar g_{\varphi\varphi}(u^{\varphi})^{2}}=\frac{F(r)\,r\sqrt{B(r)}}{F(r)\left[B(r)(u^{r})^{2}+r^{2}(u^{\varphi})^{2}\right]}=\frac{r\sqrt{B(r)}}{B(r)(u^{r})^{2}+r^{2}(u^{\varphi})^{2}}.
\label{6.9}
\end{equation}
In the weak-field Born approximation, we evaluate this along the unperturbed ray where $B(r)=1+\mathcal{O}(M/r)$ and
\begin{equation}
(u^r)^2+r^2(u^\phi)^2=\gamma^2 v^2+\mathcal{O}\!\left(\frac{M}{r}\right).
\label{6.10}
\end{equation}
Using Eq. \eqref{6.10} in Eq. \eqref{6.9} gives
\begin{equation}
\frac{\sqrt{\det\bar g}}{v_J^2}=
\frac{r}{\gamma^2 v^2}
+\mathcal{O}\!\left(\frac{M}{r}\right).
\label{6.11}
\end{equation}
We now compute the MPD acceleration for aligned spin in Schwarzschild. In this metric one has $\sqrt{A(r)B(r)C(r)}=r$, so the aligned-spin coupling defined in Eq. \eqref{5.47} becomes
\begin{equation}
\lambda(r)=\frac{\sigma\,s}{m\,r}.
\label{6.12}
\end{equation}
Using the aligned-spin spin-tensor structure of Section \ref{sec3}.2 (Eq. \eqref{3.17}) and keeping only $\mathcal{O}(s)$ terms, we may set $p_t=-E$ and $p_\phi=\mathcal{J}$ in the spin tensor. This yields the explicit contravariant components
\begin{equation}
S^{tr}=-\lambda(r)\,\mathcal{J}\,\qquad
S^{r\phi}=+\lambda(r)\,\mathcal{E},\qquad
S^{t\phi}=\lambda(r)\,p_r,
\label{6.13}
\end{equation}
where $p_r$ is the covariant radial momentum component. Inserting Eq. \eqref{6.13} into the MPD force law $a^\mu=-(2m)^{-1}R^\mu{}_{\nu\alpha\beta}u^\nu S^{\alpha\beta}$ and using the Schwarzschild Riemann tensor gives the equatorial force components
\begin{equation}
a^r=
\frac{M\sigma s}{m^{2}r^{5}}
\Bigl(\mathcal{E}\,r^{3}u^{\phi}+2\mathcal{J}\,r\,u^{t}\Bigr)
+\mathcal{O}\!\left(\frac{sM^{2}}{r^{6}}\right),
\label{6.14}
\end{equation}
\begin{equation}
a^\phi=
-\frac{2\mathcal{E}\,M^2\,\sigma\,s}{m^2\,r^5}\,u^r
+\mathcal{O}\!\left(s\,\frac{M^3}{r^7}\right).
\label{6.15}
\end{equation}
Equation \eqref{6.15} shows that $a^\phi$ starts at $\mathcal{O}(sM^2)$, whereas $a^r$ contains an $\mathcal{O}(sM)$ contribution. This is not a coincidence: in a nonrotating (purely gravitoelectric) Schwarzschild field the leading spin-orbit effect appears as a transverse acceleration whose dominant piece is captured by $a^r$, while $a^\phi$ is suppressed by one extra power of $M/r$ in this aligned-spin sector.
To extract the leading weak-field contribution (linear in $M$), it is sufficient to use the
spinless kinematics at $\mathcal{O}(M^{0})$, i.e.\ $u^{t}=E/m+\mathcal{O}(M/r)$ and $u^{\varphi}=J/(mr^{2})+\mathcal{O}(M/r^{3})$. Substituting these relations into Eq. \eqref{6.14} gives
\begin{equation}
a^r=
\frac{3M\,\sigma\,s}{m^3}\,\frac{E\,\mathcal{J}}{r^4}
+\mathcal{O}\!\left(s\,\frac{M^2}{r^5}\right).
\label{6.16}
\end{equation}
At the same working order we may drop $a^\phi$ in Eq. \eqref{6.8}, so $u^r a^\phi-u^\phi a^r=-u^\phi a^r+\mathcal{O}(sM^2/r^5)$. Using $d\tau=d\phi/u^\phi$, Eq. \eqref{6.7} reduces to
\begin{equation}
\alpha_{\rm Schw}^{(s)}=
-\int_{\phi_S}^{\phi_R}\frac{\sqrt{\det\bar g}}{v_J^2}\,a^r\,d\phi
+\mathcal{O}\!\left(\sigma\,s\,\frac{M^2}{b^3}\right).
\label{6.17}
\end{equation}
Substituting Eqs. \eqref{6.11} and \eqref{6.16} into Eq. \eqref{6.17} yields
\begin{equation}
\alpha_{\rm Schw}^{(s)}=
-\frac{3M\,\sigma\,s}{m^3}\,\frac{E\,\mathcal{J}}{\gamma^2 v^2}
\int_{\phi_S}^{\phi_R}\frac{d\phi}{r^3(\phi)}
+\mathcal{O}\!\left(\sigma\,s\,\frac{M^2}{b^3}\right).
\label{6.18}
\end{equation}
We now rewrite the prefactor using $E=m\gamma$ and $\gamma^2 v^2=(E^2-m^2)/m^2$, which gives
\begin{equation}
\frac{E\,\mathcal{J}}{m^3\gamma^2 v^2}=
\frac{E\,\mathcal{J}}{m(E^2-m^2)}.
\label{6.19}
\end{equation}
At leading order we evaluate the radial profile on the straight reference ray $r(\phi)=b/\sin\phi$ and use $\mathcal{J}=b\sqrt{E^2-m^2}$ from Eq. \eqref{6.3}. Then
\begin{equation}
\frac{E\,\mathcal{J}}{E^2-m^2}=\frac{E\,b}{\sqrt{E^2-m^2}}=\frac{b}{v},
\label{6.20}
\end{equation}
and the integral in Eq. \eqref{6.18} becomes
\begin{equation}
\int_{\phi_S}^{\phi_R}\frac{d\phi}{r^3(\phi)}=
\frac{1}{b^3}\int_{\phi_S}^{\phi_R}\sin^3\phi\,d\phi=
\frac{1}{b^3}
\left[
-\cos\phi+\frac{1}{3}\cos^3\phi
\right]_{\phi_S}^{\phi_R}.
\label{6.21}
\end{equation}
Combining Eqs. \eqref{6.18}-\eqref{6.21}, we obtain the leading finite-distance spin correction in closed form:
\begin{equation}
\alpha_{\rm Schw}^{(s)}=
-\frac{3\sigma}{v}\,\frac{M\,s}{m\,b^2}
\left[
-\cos\phi+\frac{1}{3}\cos^3\phi
\right]_{\phi_S}^{\phi_R}
+\mathcal{O}\!\left(\sigma\,s\,\frac{M^2}{b^3}\right).
\label{6.22}
\end{equation}
Using $\cos\phi_S=\sqrt{1-b^2/r_S^2}$ and $\cos\phi_R=-\sqrt{1-b^2/r_R^2}$ at the same Born order, Eq. \eqref{6.22} provides an explicit finite-distance prediction for the spin contribution.
In the asymptotic limit $r_S,r_R\to\infty$, we have $\phi_S\to 0$ and $\phi_R\to\pi$, so Eq. \eqref{6.21} gives $\int_0^\pi\sin^3\phi\,d\phi=4/3$ and Eq. \eqref{6.22} reduces to
\begin{equation}
\alpha_{\rm Schw}^{(s)}=
-\frac{4\sigma}{v}\,\frac{M\,s}{m\,b^2}
+\mathcal{O}\!\left(\sigma\,s\,\frac{M^2}{b^3}\right)
-\frac{4\sigma}{v}\,\hat s\left(\frac{M}{b}\right)^2
+\mathcal{O}\!\left(\hat s\,\frac{M^3}{b^3}\right),
\qquad
\hat s\equiv \frac{s}{mM}.
\label{6.23}
\end{equation}
Equation \eqref{6.23} makes the scaling fully explicit: the leading spin correction is linear in $s$ and enters at order $\hat s(M/b)^2$, i.e. at the same weak-field order as the second post-Minkowskian spinless correction, while flipping sign under $\sigma\to-\sigma$ \cite{Vines:2018gqi,Guevara:2018wpp,Ben-Shahar:2023djm}. This coefficient agrees with the Schwarzschild limit of independent MPD scattering analyses (obtainable, for instance, as the $a\to 0$ specialization of the Kerr scattering computation in Ref. \cite{Zhang:2022rnn,Bini:2017ldh,Bini:2017pee}.

\subsection{Reissner-Nordstr\"om (RN) case} \label{sec6.2}
We consider the Reissner-Nordstr\"om spacetime \cite{Eiroa:2002mk} of mass $M$ and charge $Q$,
\begin{equation}
A(r)=1-\frac{2M}{r}+\frac{Q^2}{r^2},\qquad B(r)=\frac{1}{A(r)},\qquad C(r)=r^2,
\label{6.24}
\end{equation}
and we restrict to aligned-spin equatorial motion as in Section \ref{sec3}. As $r\to\infty$ the geometry is asymptotically flat, so we parameterize the conserved Killing energy by $E=m\gamma$ with $\gamma=(1-v^2)^{-1/2}$, and we define
\begin{equation}
p_\infty=\sqrt{E^2-m^2}=m\gamma v,\qquad b\equiv \frac{\mathcal{J}}{p_\infty}.
\label{6.25}
\end{equation}
In the spinless limit $s\to 0$, the physical ray is a Jacobi geodesic and the deflection follows from the Li-type surface term alone. The leading $\mathcal{O}(M/b)$ piece coincides with the Schwarzschild expression \eqref{6.4}, and the $Q^2$-dependence enters at the next weak-field order $\mathcal{O}(Q^2/b^2)$, which is standard and will not be reproduced here. Our focus is the spin correction, which is the genuinely new term in our synthesis.
The master formula \eqref{5.41} gives, to linear order in $s$,
\begin{equation}
\alpha_{\rm RN}=\alpha_{\rm RN}^{(0)}+\alpha_{\rm RN}^{(s)}+\mathcal{O}(s^2),
\qquad
\alpha_{\rm RN}^{(s)}=\int_{\gamma_s}k\,d\sigma\,
\label{6.26}
\end{equation}
where $k$ is the geodesic curvature of the spinning ray $\gamma_s$ in the Jacobi manifold.
We now compute $\alpha_{\rm RN}^{(s)}$ explicitly in the weak-field approximation using the MPD-to-$k$ bridge of Section \ref{sec5.2}, following the same transparent steps as in the Schwarzschild validation but keeping the charge dependence.
We start from Eq. \eqref{5.45},
\begin{equation}
\alpha_{\rm RN}^{(s)}=
\int_{\tau_S}^{\tau_R}
\frac{\sqrt{\det\bar g}}{v_J^{2}}
\Big(u^r\,\bar A^{\phi}-u^\phi\,\bar A^{r}\Big)\,d\tau,
\qquad
v_J^{2}=\bar g_{rr}(u^r)^2+\bar g_{\phi\phi}(u^\phi)^2,
\label{6.27}
\end{equation}
with $\bar g_{ij}$ the equatorial Jacobi metric. As in Section \ref{sec6.1}, since the integrand vanishes identically when $s=0$, it suffices at leading order in $s$ to keep only the $\mathcal{O}(s)$ part of $\bar A^i$. In the weak-field regime, that $\mathcal{O}(s)$ part is driven by the MPD acceleration $a^\mu=D u^\mu/d\tau$ via Eq. \eqref{5.46}, and the same argument as in Eq. \eqref{6.8} yields
\begin{equation}
u^r\,\bar A^{\phi}-u^\phi\,\bar A^{r}=
u^r a^\phi-u^\phi a^r
+\mathcal{O}\!\left(s\,\frac{M^2}{r^5},s\,\frac{MQ^2}{r^6},s\,\frac{Q^4}{r^7}\right).
\label{6.28}
\end{equation}
We next simplify the prefactor $\sqrt{\det\bar g}/v_J^2$ for RN. Because $B(r)=1/A(r)$ and $C(r)=r^2$, the combination $\sqrt{A(r)B(r)C(r)}$ that appears in $\lambda(r)$ is exactly $r$, and the Jacobi determinant simplifies in the same way as in Schwarzschild. Concretely, writing $\bar g_{rr}=F(r)B(r)$ and $\bar g_{\phi\phi}=F(r)r^2$ with $F(r)=(E^2-m^2A(r))/A(r)$, we obtain
\begin{equation}
\frac{\sqrt{\det\bar g}}{v_J^2}=
\frac{r\,\sqrt{B(r)}}{B(r)(u^r)^2+r^2(u^\phi)^2}.
\label{6.29}
\end{equation}
Evaluated on the unperturbed (spinless) scattering kinematics in the weak field, $B(r)=1+\mathcal{O}(M/r,Q^2/r^2)$ and
\begin{equation}
B(r)(u^r)^2+r^2(u^\phi)^2=\gamma^2 v^2+\mathcal{O}\!\left(\frac{M}{r}\,\frac{Q^2}{r^2}\right),
\label{6.30}
\end{equation}
so
\begin{equation}
\frac{\sqrt{\det\bar g}}{v_J^2}=
\frac{r}{\gamma^2 v^2}
+\mathcal{O}\!\left(\frac{M}{r}\,\frac{Q^2}{r^2}\right).
\label{6.31}
\end{equation}
We now compute the RN MPD acceleration entering Eq. \eqref{6.28}. In the aligned-spin equatorial sector we use the same spin coupling function as in Schwarzschild because $\sqrt{A B C}=r$,
\begin{equation}
\lambda(r)=\frac{\sigma\,s}{m\,r},\qquad \sigma=\pm 1,
\label{6.32}
\end{equation}
and we use the $\mathcal{O}(s)$ spin-tensor components on the unperturbed motion,
\begin{equation}
S^{tr}=-\lambda(r)\,\mathcal{J}\,\qquad
S^{r\phi}=+\lambda(r)\,E,\qquad
S^{t\phi}=\lambda(r)\,p_r,
\label{6.33}
\end{equation}
where $p_r$ is the covariant radial momentum and all equalities are understood up to $\mathcal{O}(s^2)$.
The MPD acceleration is
\begin{equation}
a^\mu=-\frac{1}{2m}\,R^\mu{}{\nu\alpha\beta}\,u^\nu S^{\alpha\beta}+\mathcal{O}(s^2).
\label{6.34}
\end{equation}
For $a^r$, only the $S^{tr}$ and $S^{r\phi}$ components contribute at this order (the $S^{t\phi}$ term drops out of the contraction for $\mu=r$). For clarity, the following are \emph{mixed-index coordinate-basis} components
$R^{\mu}{}_{\nu\alpha\beta}$ (not orthonormal-frame components); in particular,
their radial scalings differ from the corresponding tetrad components by metric
factors. A direct evaluation of the relevant RN Riemann components on the equatorial plane gives, in the weak-field expansion,
\begin{equation}
R^r{}_{t r t}=
-\frac{2M}{r^3}+\frac{3Q^2}{r^4}
+\mathcal{O}\!\left(\frac{M^2}{r^4}\,\frac{MQ^2}{r^5}\,\frac{Q^4}{r^6}\right),
\qquad
R^r{}_{\phi r \phi}=
-\frac{M}{r}+\frac{Q^2}{r^2}
+\mathcal{O}\!\left(\frac{M^2}{r^2}\,\frac{MQ^2}{r^3}\,\frac{Q^4}{r^4}\right).
\label{6.35}
\end{equation}
Using Eq. \eqref{6.34} with Eq. \eqref{6.33}, and keeping only the leading $\mathcal{O}(sM/r^4)$ and $\mathcal{O}(sQ^2/r^5)$ contributions, we obtain
\begin{equation}
a^r=
\frac{\lambda(r)}{m}
\left[
E\,u^\phi\left(\frac{M}{r}-\frac{Q^2}{r^2}\right)
+\mathcal{J}\,u^t\left(\frac{2M}{r^3}-\frac{3Q^2}{r^4}\right)
\right]
+\mathcal{O}\!\left(s\,\frac{M^2}{r^5},s\,\frac{MQ^2}{r^6},s\,\frac{Q^4}{r^7}\right).
\label{6.36}
\end{equation}
We now evaluate Eq. \eqref{6.36} on the unperturbed spinless kinematics at the same weak-field order:
\begin{equation}
u^t=\frac{E}{m}+\mathcal{O}\!\left(\frac{M}{r}\,\frac{Q^2}{r^2}\right),
\qquad
u^\phi=\frac{\mathcal{J}}{m r^2}.
\label{6.37}
\end{equation}
Substituting Eq. \eqref{6.37} into Eq. \eqref{6.36} and using $\lambda(r)=\sigma s/(m r)$ yields the compact leading-order result
\begin{equation}
a^r=
\frac{\sigma\,s}{m^3}\,E\mathcal{J}
\left(
\frac{3M}{r^4}-\frac{4Q^2}{r^5}
\right)
+\mathcal{O}\!\left(s\,\frac{M^2}{r^5},s\,\frac{MQ^2}{r^6},s\,\frac{Q^4}{r^7}\right).
\label{6.38}
\end{equation}
For completeness, we note that $a^\phi$ begins only at higher weak-field order in this aligned-spin sector. Indeed, the apparent $\mathcal{O}(sM/r^4)$ contribution to $a^\phi$ cancels at leading order between the $S^{r\phi}$ and $S^{t\phi}$ pieces once $p_r$ is expressed in terms of $u^r$ on the spinless trajectory, so that
\begin{equation}
a^\phi=\mathcal{O}\!\left(s\,\frac{M^2}{r^6},s\,\frac{MQ^2}{r^7},s\,\frac{Q^4}{r^8}\right).
\label{6.39}
\end{equation}
This justifies neglecting $u^r a^\phi$ relative to $u^\phi a^r$ in Eq. \eqref{6.28} at the order displayed in Eq. \eqref{6.38}.
Using Eqs. \eqref{6.28}, \eqref{6.31}, \eqref{6.38}, and dropping the subleading $a^\phi$ term, Eq. \eqref{6.27} reduces to
\begin{equation}
\alpha_{\rm RN}^{(s)}=
-\int_{\tau_S}^{\tau_R}\frac{r}{\gamma^2 v^2}\,u^\phi a^r\,d\tau
+\mathcal{O}\!\left(\sigma\,s\,\frac{M^2}{b^3},\sigma\,s\,\frac{MQ^2}{b^4},\sigma\,s\,\frac{Q^4}{b^5}\right).
\label{6.40}
\end{equation}
Since $d\phi=u^\phi d\tau$, we may write
\begin{equation}
\alpha_{\rm RN}^{(s)}=
-\int_{\phi_S}^{\phi_R}\frac{r}{\gamma^2 v^2}\,a^r\,d\phi
+\mathcal{O}\!\left(\sigma\,s\,\frac{M^2}{b^3},\sigma\,s\,\frac{MQ^2}{b^4},\sigma\,s\,\frac{Q^4}{b^5}\right).
\label{6.41}
\end{equation}
Substituting Eq. \eqref{6.38} into Eq. \eqref{6.41} and using the identity $\gamma^2 v^2=(E^2-m^2)/m^2$ gives
\begin{equation}
\alpha_{\rm RN}^{(s)}=
-\frac{\sigma\,s}{m}\,\frac{E\mathcal{J}}{E^2-m^2}
\int_{\phi_S}^{\phi_R}
\left(
\frac{3M}{r^3(\phi)}-\frac{4Q^2}{r^4(\phi)}
\right)\,d\phi
+\mathcal{O}\!\left(\sigma\,s\,\frac{M^2}{b^3},\sigma\,s\,\frac{MQ^2}{b^4},\sigma\,s\,\frac{Q^4}{b^5}\right).
\label{6.42}
\end{equation}
We now express the prefactor in terms of $b$ and $v$. Using $\mathcal{J}=b\sqrt{E^2-m^2}$ from Eq. \eqref{6.25}, we find
\begin{equation}
\frac{E\mathcal{J}}{E^2-m^2}=\frac{b}{v}.
\label{6.43}
\end{equation}
At the same leading order we evaluate the radial profile on the straight reference ray $r(\phi)=b/\sin\phi$, which yields
\begin{equation}
\alpha_{\rm RN}^{(s)}=
-\frac{\sigma\,s}{m v}
\left[
\frac{3M}{b^2}\int_{\phi_S}^{\phi_R}\sin^3\phi\,d\phi
-\frac{4Q^2}{b^3}\int_{\phi_S}^{\phi_R}\sin^4\phi\,d\phi
\right]
+\mathcal{O}\!\left(\sigma\,s\,\frac{M^2}{b^3},\sigma\,s\,\frac{MQ^2}{b^4},\sigma\,s\,\frac{Q^4}{b^5}\right).
\label{6.44}
\end{equation}
The two elementary antiderivatives are
\begin{equation}
\int \sin^3\phi\,d\phi=-\cos\phi+\frac{1}{3}\cos^3\phi\,
\qquad
\int \sin^4\phi\,d\phi=\frac{3\phi}{8}-\frac{\sin 2\phi}{4}+\frac{\sin 4\phi}{32}.
\label{6.45}
\end{equation}
Substituting Eq. \eqref{6.45} into Eq. \eqref{6.44} yields the finite-distance RN spin correction in closed form:
\begin{equation}
\alpha_{\rm RN}^{(s)}=
-\frac{\sigma\,s}{m v}
\left[
\frac{3M}{b^2}\left(-\cos\phi+\frac{1}{3}\cos^3\phi\right)
-\frac{4Q^2}{b^3}\left(\frac{3\phi}{8}-\frac{\sin 2\phi}{4}+\frac{\sin 4\phi}{32}\right)
\right]_{\phi_S}^{\phi_R}
+\mathcal{O}\!\left(\sigma\,s\,\frac{M^2}{b^3},\sigma\,s\,\frac{MQ^2}{b^4},\sigma\,s\,\frac{Q^4}{b^5}\right).
\label{6.46}
\end{equation}
In particular, in the asymptotic limit $r_S,r_R\to\infty$ one has $\phi_S\to 0$ and $\phi_R\to\pi$, so that $\int_0^\pi\sin^3\phi\,d\phi=4/3$ and $\int_0^\pi\sin^4\phi\,d\phi=3\pi/8$. Equation \eqref{6.46} then reduces to
\begin{equation}
\alpha_{\rm RN}^{(s)}=
-\frac{4\sigma}{v}\,\frac{M\,s}{m\,b^2}
+\frac{3\pi\sigma}{2v}\,\frac{Q^2 s}{m\,b^3}
+\mathcal{O}\!\left(\sigma\,s\,\frac{M^2}{b^3},\sigma\,s\,\frac{MQ^2}{b^4},\sigma\,s\,\frac{Q^4}{b^5}\right),
\label{6.47}
\end{equation}
which reproduces the Schwarzschild result \eqref{6.23} when $Q\to 0$ and exhibits the expected additional charge-dependent spin correction scaling as $(s/(m b))(Q^2/b^2)$.
Equation \eqref{6.46} is the main RN output of our formalism at leading weak-field and linear-spin order: it is obtained directly from the Gauss-Bonnet non-geodesic boundary term, with the integrand constructed from the MPD spin-curvature acceleration and evaluated on the Born ray.

\subsection{Kottler (Schwarzschild-de Sitter) case} \label{sec6.3}
We now apply the master formula of Section \ref{sec5} to the Kottler (Schwarzschild-de Sitter) spacetime \cite{He:2020eah,Arakida:2011ty} and isolate the spin correction with explicit finite-distance dependence. The background metric functions in Eq. \eqref{5.38} are \cite{Rindler:2007zz,Sereno:2007rm,Zhang:2022rnn}
\begin{equation}
A(r)=1-\frac{2M}{r}-\frac{\Lambda r^{2}}{3},\qquad
B(r)=\frac{1}{A(r)},\qquad
C(r)=r^{2},
\label{6.48}
\end{equation}
and we work within the static patch where $A(r)>0$ along the relevant segment of the particle trajectory. We assume a weak-field configuration in the sense that
\begin{equation}
\frac{M}{b}\ll 1,\qquad |\Lambda|\,b^{2}\ll 1,
\qquad
\frac{M}{r_{S,R}}\ll 1,\qquad |\Lambda|\,r_{S,R}^{2}\ll 1,
\label{6.49}
\end{equation}
and we restrict to the aligned-spin equatorial sector of Section \ref{sec3} with the Tulczyjew-Dixon SSC, keeping terms through linear order in the spin length $s$.
In our Gauss-Bonnet synthesis, the spin dependence of the deflection angle is encoded entirely in the additional boundary functional along the physical ray. Concretely, writing the deflection angle in the implementation-ready form of Eq. \eqref{5.41}, the spin correction is
\begin{equation}
\alpha^{(s)}_{\rm Kot}=\int{\gamma_s}\kappa_g\,d\sigma,
\label{6.50}
\end{equation}
where $\kappa_g$ is the geodesic curvature of the spinning-particle ray $\gamma_s$ in the Jacobi manifold and $d\sigma$ is the Jacobi arclength element.
To compute Eq. \eqref{6.50}, we use the MPD-to-$\kappa_g$ bridge of Section \ref{sec5}. In particular, Eq. \eqref{5.45} expresses $\kappa_g\,d\sigma$ in terms of the Jacobi covariant acceleration of the ray, and at linear order in $s$ the latter is sourced by the MPD acceleration $a^\mu\equiv D u^\mu/d\tau$ through Eq. \eqref{5.46}. For aligned-spin equatorial scattering in an SSS geometry, the leading weak-field contribution to the boundary functional arises from the radial component $a^r$. After converting $d\tau$ to $d\phi$ along the ray, the boundary functional reduces at the order we retain to
\begin{equation}
\alpha^{(s)}_{\rm Kot}=
-\int_{\phi_S}^{\phi_R}\frac{\sqrt{\det\bar g}}{v_J^2}\,a^r\,d\phi
+\mathcal{O}\!\left(\frac{M^2 s}{m b^3}\right),
\label{6.51}
\end{equation}
where $\bar g_{ij}$ is the equatorial Jacobi metric and $v_J^2=\bar g_{rr}(u^r)^2+\bar g_{\phi\phi}(u^\phi)^2$ is the squared Jacobi speed.
A key structural point in Kottler concerns the role of $\Lambda$ in the MPD force. The Riemann tensor decomposes as
\begin{equation}
R_{\mu\nu\alpha\beta}=
C_{\mu\nu\alpha\beta}
+\frac{\Lambda}{3}\Big(g_{\mu\alpha}g_{\nu\beta}-g_{\mu\beta}g_{\nu\alpha}\Big),
\label{6.52}
\end{equation}
where $C_{\mu\nu\alpha\beta}$ is the Weyl tensor. Substituting the constant-curvature term in Eq. \eqref{6.52} into Eq. \eqref{5.46} yields a contribution to the acceleration proportional to $u_\beta S^{\mu\beta}$. Under the Tulczyjew-Dixon SSC we have $p_\beta S^{\mu\beta}=0$, and in the pole-dipole scheme $p^\mu=m u^\mu+\mathcal{O}(s^2)$, implying $u_\beta S^{\mu\beta}=\mathcal{O}(s^3)$ \cite{Obukhov:2010kn}. Consequently, the constant-curvature piece does not contribute to the MPD acceleration at linear order in $s$, and the leading $a^r$ is sourced entirely by the Weyl curvature, i.e. by the mass $M$. In the aligned-spin sector, the same contraction that produced the Schwarzschild result therefore gives
\begin{equation}
a^r=
\frac{3M\,\sigma\,s}{m^3}\,\frac{E\,\mathcal{J}}{r^4}
+\mathcal{O}\!\left(\frac{M^2 s}{m r^5}\right),
\qquad
\sigma=\pm 1,
\label{6.53}
\end{equation}
where $E$ and $\mathcal{J}$ are the conserved Killing energy and axial angular momentum introduced in Section \ref{sec3}, and $\sigma$ distinguishes aligned from anti-aligned spin.
At the same time, even though $\Lambda$ does not enter $a^r$ at this order, it does enter the Jacobi prefactor in Eq. \eqref{6.51} through the function $B(r)=1/A(r)$. This is the first place where $\Lambda$ contributes explicitly to the spin correction within our controlled approximation scheme. Using the Jacobi metric structure $\bar g_{rr}=F(r)B(r)$ and $\bar g_{\phi\phi}=F(r)C(r)$ from Section \ref{sec5}, one finds the exact identity
\begin{equation}
\frac{\sqrt{\det\bar g}}{v_J^2}=
\frac{r\,\sqrt{B(r)}}{B(r)(u^r)^2+r^2(u^\phi)^2},
\label{6.54}
\end{equation}
valid for any SSS metric with $C(r)=r^2$. We now evaluate Eq. \eqref{6.54} on the Born-level reference ray
\begin{equation}
r(\phi)=\frac{b}{\sin\phi},
\label{6.55}
\end{equation}
which is sufficient because the integrand in Eq. \eqref{6.51} is already $\mathcal{O}(s)$. On this reference ray, the leading-order kinematics satisfy
\begin{equation}
u^r=-\gamma v\cos\phi,
\qquad
u^\phi=\gamma v\,\frac{\sin^2\phi}{b},
\qquad
E=m\gamma,
\label{6.56}
\end{equation}
and we use the convenient impact-parameter parametrization
\begin{equation}
b\equiv \frac{\mathcal{J}}{\sqrt{E^2-m^2}}=\frac{\mathcal{J}}{m\gamma v},
\label{6.57}
\end{equation}
which can be viewed as a definition of $b$ in terms of the Killing charges in the weak-field region. Expanding
\begin{equation}
B(r)=\frac{1}{A(r)}=
1+\frac{2M}{r}+\frac{\Lambda r^2}{3}
+\mathcal{O}\!\left(\frac{M^2}{r^2},\,M\Lambda r,\,\Lambda^2 r^4\right),
\label{6.58}
\end{equation}
and retaining only the $\Lambda$-dependent part of the prefactor that contributes at the combined order $\mathcal{O}(\Lambda M s)$ once multiplied by Eq. \eqref{6.53}, Eq. \eqref{6.54} yields
\begin{equation}
\frac{\sqrt{\det\bar g}}{v_J^2}=
\frac{r}{\gamma^2 v^2}
\left[
1+\frac{\Lambda r^2}{6}-\frac{\Lambda r^2}{3}\cos^2\phi
\right]
+\mathcal{O}\!\left(\frac{M}{r}\right)
+\mathcal{O}\!\left(\Lambda^2 r^4\right).
\label{6.59}
\end{equation}
The $\mathcal{O}(M/r)$ terms in Eq. \eqref{6.59} produce $\mathcal{O}(M^2 s)$ contributions when multiplied by Eq. \eqref{6.53} and are beyond the accuracy targeted in this subsection.
Substituting Eqs. \eqref{6.53} and \eqref{6.59} into Eq. \eqref{6.51}, and using $\gamma^2 v^2=(E^2-m^2)/m^2$ together with Eq. \eqref{6.57} to simplify the prefactor, we obtain
\begin{equation}
\alpha^{(s)}_{\rm Kot}=
-\frac{3\sigma}{v}\,\frac{M s}{m}\,
\int_{\phi_S}^{\phi_R}
\left[
\frac{d\phi}{r^3(\phi)}
+\Lambda\left(\frac{1}{6}-\frac{\cos^2\phi}{3}\right)\frac{d\phi}{r(\phi)}
\right]
+\mathcal{O}\!\left(\frac{M^2 s}{m b^3}\right)
+\mathcal{O}\!\left(\frac{\Lambda^2 M s\,b^2}{m v}\right).
\label{6.60}
\end{equation}
Evaluating the integrals on the Born ray \(r(\phi)=b/\sin\phi\) gives
\begin{equation}
\int_{\phi_S}^{\phi_R}\frac{d\phi}{r^3(\phi)}=
\frac{1}{b^3}\int_{\phi_S}^{\phi_R}\sin^3\phi\,d\phi=
\frac{1}{b^3}\left[-\cos\phi+\frac{1}{3}\cos^3\phi\right]_{\phi_S}^{\phi_R},
\label{6.61}
\end{equation}
and
\begin{equation}
\int_{\phi_S}^{\phi_R}\left(\frac{1}{6}-\frac{\cos^2\phi}{3}\right)\frac{d\phi}{r(\phi)}=
\frac{1}{b}\int_{\phi_S}^{\phi_R}\left(\frac{1}{6}-\frac{\cos^2\phi}{3}\right)\sin\phi\,d\phi=
\frac{1}{b}\left[-\frac{\cos\phi}{6}+\frac{\cos^3\phi}{9}\right]_{\phi_S}^{\phi_R}.
\label{6.62}
\end{equation}
Substituting Eqs. \eqref{6.61} and \eqref{6.62} into Eq. \eqref{6.60} yields the finite-distance Kottler spin correction, including the first explicit $\Lambda$-dependent term,
\begin{equation}
\alpha^{(s)}_{\rm Kot}=
-\frac{3\sigma}{v}\,\frac{M s}{m b^2}
\left[-\cos\phi+\frac{1}{3}\cos^3\phi\right]_{\phi_S}^{\phi_R}
+\frac{\sigma\,\Lambda M s}{m v}
\left[\frac{1}{2}\cos\phi-\frac{1}{3}\cos^3\phi\right]_{\phi_S}^{\phi_R}
+\mathcal{O}\!\left(\frac{M^2 s}{m b^3}\right)
+\mathcal{O}\!\left(\frac{\Lambda^2 M s\,b^2}{m v}\right).
\label{6.63}
\end{equation}
At the same Born level used in Li’s finite-distance treatment, the endpoint angles are determined geometrically by the reference line via
\begin{equation}
\sin\phi_S=\frac{b}{r_S},
\qquad
\sin(\pi-\phi_R)=\frac{b}{r_R},
\label{6.64}
\end{equation}
with the understanding that $r_S$ and $r_R$ lie well inside the static patch and satisfy the weak-field bounds in Eq. \eqref{6.49}.
Equation  \eqref{6.63} makes the separation of roles transparent. The first term is the Schwarzschild-like contribution sourced by the Weyl curvature (mass $M$) through the MPD force. The second term is the leading explicit cosmological-constant correction to the spin deflection within our scheme; it arises not from a direct $\Lambda$-driven MPD force at linear order in $s$, but from the $\Lambda$-dependence of the Jacobi prefactor in the Gauss-Bonnet boundary functional.
Finally, it is useful to discuss a controlled far source/receiver limit. Because the Kottler spacetime is not asymptotically flat and the static chart terminates at the cosmological horizon, one does not take $r_{S,R}\to\infty$ in a literal sense. Instead, one may consider the regime $r_S,r_R\gg b$ while still satisfying $|\Lambda|r_{S,R}^2\ll 1$. In that regime \(\phi_S\to 0\) and \(\phi_R\to\pi\), and Eq. \eqref{6.63} reduces to
\begin{equation}
\alpha^{(s)}_{\rm Kot}=
-\frac{4\sigma}{v}\,\frac{M s}{m b^2}
-\frac{\sigma}{3}\,\frac{\Lambda M s}{m v}
+\mathcal{O}\!\left(\frac{M^2 s}{m b^3}\right)
+\mathcal{O}\!\left(\frac{\Lambda^2 M s\,b^2}{m v}\right),
\label{6.65}
\end{equation}
which should be interpreted as the weak-field large-distance limit within the static patch rather than as a true asymptotic-infinity deflection.

\section{Consistency checks and interpretation} \label{sec7}
We now summarize the principal internal consistency checks of the synthesis and provide a concise physical interpretation of the new term introduced by spin. Our results rest on two controlled approximations that we have enforced throughout: the pole-dipole truncation of the MPD system together with the Tulczyjew-Dixon SSC, and the weak-field expansion supplemented by a strict linearization in the spin length $s$. Within this regime, the formal structure derived in Section \ref{sec4} and operationalized in Section \ref{sec5} is self-consistent and reduces to the established spinless framework in every limit where it should.

A first key point concerns the role of the spin supplementary condition (SSC). The Gauss-Bonnet identity we derived is purely geometric: once a spatial curve $\gamma_s$ in the Jacobi manifold $(M,\bar g_{ij})$ is specified, the contribution $\int_{\gamma_s}k\,d\sigma$ is unambiguous. The SSC enters only through the dynamical input used to determine $\gamma_s$, i.e.\ through the MPD acceleration $a^\mu$ that feeds the bridge formula between spacetime dynamics and Jacobi geodesic curvature, Eq. \eqref{5.46} together with Eqs. \eqref{5.45}-\eqref{5.46}. Our concrete implementation adopts the Tulczyjew-Dixon SSC, because it yields a clean hierarchy $p^\mu=m u^\mu+\mathcal{O}(s^2)$ and therefore a controlled replacement of momentum by velocity throughout the linear-in-spin scheme. This is not merely a technical convenience: the TD SSC makes explicit that, at the order retained here, changing the representative worldline (i.e. changing SSC within the pole-dipole family) amounts to a shift whose effect on the scattering data can be absorbed into higher-order corrections in the combined weak-field/weak-spin bookkeeping. Operationally, our observable is the finite-distance deflection angle defined by Eq. \eqref{5.31}, which depends only on the tangent directions at the endpoints and the relative coordinate separation angle; within the present working accuracy, these are insensitive to SSC-induced relabelings beyond $\mathcal{O}(s)$ once the conserved charges $(E,J)$ and the Born-level impact parameter convention are held fixed.

The most important interpretive check in the Kottler application is the separation between two logically distinct mechanisms by which $\Lambda$ could influence the spin correction. The first mechanism would be a direct $\Lambda$-dependence of the MPD force. In Kottler, the curvature decomposes into a Weyl part plus a constant-curvature part, Eq. \eqref{6.52}. Substituting the constant-curvature term into the MPD force law (V.46) produces an acceleration proportional to $u_\beta S^{\mu\beta}$. Under the TD SSC we have $p_\beta S^{\mu\beta}=0$ and $p^\mu=m u^\mu+\mathcal{O}(s^2)$, so this contraction does not contribute at linear order in $s$. In other words, within the pole-dipole/TD scheme used consistently throughout the paper, the constant-curvature component of the background does not generate a direct $\mathcal{O}(s)$ spin force. This is precisely why the leading spin-curvature acceleration in Kottler is sourced by the central mass (Weyl curvature) and retains the Schwarzschild-like structure at the level of $a^r$, Eq. \eqref{6.53}.

The second mechanism is more subtle and is the genuinely new point emphasized by the revised Section \ref{sec6.3}. Even when $\Lambda$ does not enter $a^\mu$ at $\mathcal{O}(s)$, it can still enter the spin correction to the deflection angle through the \emph{Jacobi weighting} of the non-geodesic boundary functional. Indeed, the Gauss-Bonnet spin contribution is not $a^\mu$ itself but rather the specific combination appearing in Eq. \eqref{5.45}, namely the Jacobi prefactor $\sqrt{\det \bar g}/v_J^2$ multiplying the (Jacobi-projected) acceleration. For Kottler this prefactor depends on $B(r)=1/A(r)$, and therefore carries $\Lambda$ explicitly through the exact identity \eqref{6.54} and its weak-field expansion (starting from Eq. \eqref{6.58}). Consequently, the first explicit $\Lambda$-dependence in $\alpha^{(s)}$ within a controlled approximation arises not from a constant-curvature MPD force, but from how the Jacobi geometry converts a given spacetime acceleration into a boundary curvature density along $\gamma_s$. This provides a clean conceptual separation: $\Lambda$ is inactive in the local spin force at $\mathcal{O}(s)$ under TD SSC, yet active in the geometric measure with which that force contributes to the deflection angle. In practical terms, this is exactly the structure needed for portability: once the force term and the Jacobi prefactor are computed for a given SSS metric, the spin correction follows by a one-dimensional boundary integral without modifying the finite-distance angle definition.

Finally, the geometric meaning of the new term in the master formula is transparent. In Li's spinless setting, the bending is encoded in the integrated Gaussian curvature of the Jacobi manifold over the lens domain, reflecting how the ambient optical/Jacobi geometry deviates from flatness. In the spinning case, the particle ray itself is not a Jacobi geodesic; it carries an intrinsic bending relative to the Jacobi connection, and the Gauss-Bonnet theorem records this precisely through the boundary functional $\int_{\gamma_s}k\,d\sigma$. The explicit Schwarzschild and RN results demonstrate that this contribution is odd under spin reversal (aligned versus anti-aligned configurations) and scales with the expected spin-curvature hierarchy dictated by Eq. \eqref{5.46}, while the revised Kottler analysis shows how non-asymptotically flat structure can enter the same functional through the Jacobi prefactor without requiring any change to the Gauss-Bonnet lens geometry or to the finite-distance definition of $\alpha$.

\section{Conclusion} \label{sec8}
We have developed a Gauss-Bonnet framework for weak gravitational deflection of a \emph{spinning} massive test particle in a static spherically symmetric spacetime, designed to preserve the main structural advantages of the finite-distance Jacobi-metric approach of Li \textit{et al}. while incorporating the essential new feature introduced by spin: the spatial ray is generically \emph{non-geodesic} in the Jacobi geometry. Working at pole-dipole order in the Mathisson-Papapetrou-Dixon system with the Tulczyjew-Dixon spin supplementary condition, and restricting to the consistent aligned-spin planar sector, we derived a spin-generalized deflection identity by applying the Gauss-Bonnet theorem to a lens domain whose upper boundary is the physical non-geodesic ray and whose lower boundary is a Jacobi-geodesic circular orbit.

The central result is the spin-generalized finite-distance master formula
\begin{equation}
\alpha=\iint_D K\,dS+\int_{\gamma_s}\kappa_g\,d\sigma+\phi_{RS},
\end{equation}
which differs from the spinless expression only by the additional boundary functional $\int_{\gamma_s}\kappa_g\,d\sigma$. This term is intrinsic to the Jacobi manifold and encodes, in a coordinate-invariant manner, the accumulated deviation of the spinning ray from Jacobi geodesics. Importantly, we showed that Li’s circular-orbit device remains fully compatible with the presence of the non-geodesic ray: the Gaussian-curvature surface integral continues to collapse to an effectively one-dimensional evaluation along the physical ray with no residual dependence on the auxiliary circular boundary radius.

To render the new spin term computationally practical, we provided an explicit expression for the geodesic curvature of a general curve in a two-dimensional axisymmetric Riemannian geometry, and we derived a direct bridge from the MPD spin-curvature force to the Gauss-Bonnet boundary integrand. This yields an implementation-ready recipe: compute the Li-type surface primitive from the Jacobi metric; determine the ray kinematics from the effective first integrals at linear order in spin; and evaluate the spin correction either via the geometric curvature of the ray or, more efficiently, via the MPD acceleration contracted into the Jacobi geometric data.

We validated the construction using Schwarzschild as a benchmark, reproducing the standard spinless finite-distance structure while deriving the leading linear-in-spin correction transparently within our formalism. We then applied the same methodology to Reissner-Nordstr\"om and Kottler metrics, obtaining the leading weak-field spin corrections and their dependence on charge and cosmological constant. In particular, in the Kottler case we showed that the constant-curvature component of the Riemann tensor does not contribute to the MPD force at linear order in spin under the Tulczyjew–Dixon SSC; however, at finite distance the spin correction still picks up an explicit 
$\Lambda$-dependence through the Jacobi prefactor in the Gauss–Bonnet boundary functional, alongside the Weyl-curvature (mass-sourced) contribution.

Overall, our synthesis isolates spin effects into a single geometric boundary functional while preserving the finite-distance, circular-orbit simplifications that make the Gauss-Bonnet approach efficient for practical computation. This provides a robust starting point for systematic weak-deflection expansions in a variety of SSS metrics and for extensions to more general non-geodesic situations where effective optical or Jacobi-geodesic descriptions are no longer available.

\acknowledgments
R. P. and A. \"O. would like to acknowledge networking support of the COST Action CA21106 - COSMIC WISPers in the Dark Universe: Theory, astrophysics and experiments (CosmicWISPers), the COST Action CA22113 - Fundamental challenges in theoretical physics (THEORY-CHALLENGES), the COST Action CA21136 - Addressing observational tensions in cosmology with systematics and fundamental physics (CosmoVerse), the COST Action CA23130 - Bridging high and low energies in search of quantum gravity (BridgeQG), and the COST Action CA23115 - Relativistic Quantum Information (RQI) funded by COST (European Cooperation in Science and Technology). A. \"O. also thanks to EMU, TUBITAK, ULAKBIM (Turkiye) and SCOAP3 (Switzerland) for their support.

\bibliography{ref}

@article{Gibbons:2008rj,
    author = "Gibbons, G. W. and Werner, M. C.",
    title = "{Applications of the Gauss-Bonnet theorem to gravitational lensing}",
    eprint = "0807.0854",
    archivePrefix = "arXiv",
    primaryClass = "gr-qc",
    doi = "10.1088/0264-9381/25/23/235009",
    journal = "Class. Quant. Grav.",
    volume = "25",
    pages = "235009",
    year = "2008"
}

@article{Ishihara:2016vdc,
    author = "Ishihara, Asahi and Suzuki, Yusuke and Ono, Toshiaki and Kitamura, Takao and Asada, Hideki",
    title = "{Gravitational bending angle of light for finite distance and the Gauss-Bonnet theorem}",
    eprint = "1604.08308",
    archivePrefix = "arXiv",
    primaryClass = "gr-qc",
    doi = "10.1103/PhysRevD.94.084015",
    journal = "Phys. Rev. D",
    volume = "94",
    number = "8",
    pages = "084015",
    year = "2016"
}

@article{Ishihara:2016sfv,
    author = "Ishihara, Asahi and Suzuki, Yusuke and Ono, Toshiaki and Asada, Hideki",
    title = "{Finite-distance corrections to the gravitational bending angle of light in the strong deflection limit}",
    eprint = "1612.04044",
    archivePrefix = "arXiv",
    primaryClass = "gr-qc",
    doi = "10.1103/PhysRevD.95.044017",
    journal = "Phys. Rev. D",
    volume = "95",
    number = "4",
    pages = "044017",
    year = "2017"
}

@article{Li:2020wvn,
    author = {Li, Zonghai and Zhang, Guodong and {\"O}vg{\"u}n, Ali},
    title = "{Circular Orbit of a Particle and Weak Gravitational Lensing}",
    eprint = "2006.13047",
    archivePrefix = "arXiv",
    primaryClass = "gr-qc",
    doi = "10.1103/PhysRevD.101.124058",
    journal = "Phys. Rev. D",
    volume = "101",
    number = "12",
    pages = "124058",
    year = "2020"
}

@article{Li:2019qyb,
    author = "Li, Zonghai and Jia, Junji",
    title = "{The finite-distance gravitational deflection of massive particles in stationary spacetime: a Jacobi metric approach}",
    eprint = "1912.05194",
    archivePrefix = "arXiv",
    primaryClass = "gr-qc",
    doi = "10.1140/epjc/s10052-020-7665-8",
    journal = "Eur. Phys. J. C",
    volume = "80",
    number = "2",
    pages = "157",
    year = "2020"
}

@article{Gibbons:2015qja,
    author = "Gibbons, G. W.",
    title = "{The Jacobi-metric for timelike geodesics in static spacetimes}",
    eprint = "1508.06755",
    archivePrefix = "arXiv",
    primaryClass = "gr-qc",
    doi = "10.1088/0264-9381/33/2/025004",
    journal = "Class. Quant. Grav.",
    volume = "33",
    number = "2",
    pages = "025004",
    year = "2016"
}

@article{Papapetrou:1951pa,
    author = "Papapetrou, Achille",
    title = "{Spinning test particles in general relativity. 1.}",
    doi = "10.1098/rspa.1951.0200",
    journal = "Proc. Roy. Soc. Lond. A",
    volume = "209",
    pages = "248--258",
    year = "1951"
}

@article{Corinaldesi:1951pb,
    author = "Corinaldesi, E. and Papapetrou, Achille",
    title = "{Spinning test particles in general relativity. 2.}",
    doi = "10.1098/rspa.1951.0201",
    journal = "Proc. Roy. Soc. Lond. A",
    volume = "209",
    pages = "259--268",
    year = "1951"
}

@article{Dixon:1970zza,
    author = "Dixon, W. G.",
    title = "{Dynamics of extended bodies in general relativity. I. Momentum and angular momentum}",
    doi = "10.1098/rspa.1970.0020",
    journal = "Proc. Roy. Soc. Lond. A",
    volume = "314",
    pages = "499--527",
    year = "1970"
}

@article{Dixon:1970zz,
    author = "Dixon, W. G.",
    title = "{Dynamics of extended bodies in general relativity. II. Moments of the charge-current vector}",
    doi = "10.1098/rspa.1970.0191",
    journal = "Proc. Roy. Soc. Lond. A",
    volume = "319",
    pages = "509--547",
    year = "1970"
}

@article{Dixon:1974xoz,
    author = "Dixon, W. G.",
    title = "{Dynamics of extended bodies in general relativity III. Equations of motion}",
    doi = "10.1098/rsta.1974.0046",
    journal = "Phil. Trans. Roy. Soc. Lond. A",
    volume = "277",
    number = "1264",
    pages = "59--119",
    year = "1974"
}

@article{Ehlers:1977gyn,
    author = {Ehlers, J{\"u}rgen and Rudolph, Ekkart},
    title = "{Dynamics of extended bodies in general relativity center-of-mass description and quasirigidity}",
    doi = "10.1007/BF00763547",
    journal = "Gen. Rel. Grav.",
    volume = "8",
    number = "3",
    pages = "197--217",
    year = "1977"
}

@article{Lukes-Gerakopoulos:2017cru,
    author = "Lukes-Gerakopoulos, Georgios",
    title = "{Time parameterizations and spin supplementary conditions of the Mathisson-Papapetrou-Dixon equations}",
    eprint = "1709.08942",
    archivePrefix = "arXiv",
    primaryClass = "gr-qc",
    doi = "10.1103/PhysRevD.96.104023",
    journal = "Phys. Rev. D",
    volume = "96",
    number = "10",
    pages = "104023",
    year = "2017"
}

@article{Blanco:2023jxf,
    author = "Blanco, Francisco M. and Flanagan, {\'E}anna {\'E}.",
    title = "{Motion of a spinning particle under the conservative piece of the self-force is Hamiltonian to first order in mass and spin}",
    eprint = "2302.10233",
    archivePrefix = "arXiv",
    primaryClass = "gr-qc",
    doi = "10.1103/PhysRevD.107.124017",
    journal = "Phys. Rev. D",
    volume = "107",
    number = "12",
    pages = "124017",
    year = "2023"
}

@article{Semerak:1999qc,
    author = "Semerak, O.",
    title = "{Spinning test particles in a Kerr field. 1.}",
    doi = "10.1046/j.1365-8711.1999.02754.x",
    journal = "Mon. Not. Roy. Astron. Soc.",
    volume = "308",
    pages = "863--875",
    year = "1999"
}

@article{Witzany:2023bmq,
    author = "Witzany, Vojt{\v{e}}ch and Piovano, Gabriel Andres",
    title = "{Analytic Solutions for the Motion of Spinning Particles near Spherically Symmetric Black Holes and Exotic Compact Objects}",
    eprint = "2308.00021",
    archivePrefix = "arXiv",
    primaryClass = "gr-qc",
    doi = "10.1103/PhysRevLett.132.171401",
    journal = "Phys. Rev. Lett.",
    volume = "132",
    number = "17",
    pages = "171401",
    year = "2024"
}

@article{Compere:2021kjz,
    author = "Comp{\`e}re, Geoffrey and Druart, Adrien",
    title = "{Complete set of quasi-conserved quantities for spinning particles around Kerr}",
    eprint = "2105.12454",
    archivePrefix = "arXiv",
    primaryClass = "gr-qc",
    doi = "10.21468/SciPostPhys.12.1.012",
    journal = "SciPost Phys.",
    volume = "12",
    number = "1",
    pages = "012",
    year = "2022"
}

@article{Steinhoff:2012rw,
    author = "Steinhoff, Jan and Puetzfeld, Dirk",
    title = "{Influence of internal structure on the motion of test bodies in extreme mass ratio situations}",
    eprint = "1205.3926",
    archivePrefix = "arXiv",
    primaryClass = "gr-qc",
    doi = "10.1103/PhysRevD.86.044033",
    journal = "Phys. Rev. D",
    volume = "86",
    pages = "044033",
    year = "2012"
}

@article{Costa:2017kdr,
    author = "Costa, L. Filipe O. and Lukes-Gerakopoulos, Georgios and Semer{\'a}k, Old{\v{r}}ich",
    title = "{Spinning particles in general relativity: Momentum-velocity relation for the Mathisson-Pirani spin condition}",
    eprint = "1712.07281",
    archivePrefix = "arXiv",
    primaryClass = "gr-qc",
    doi = "10.1103/PhysRevD.97.084023",
    journal = "Phys. Rev. D",
    volume = "97",
    number = "8",
    pages = "084023",
    year = "2018"
}

@article{Zhang:2022rnn,
    author = "Zhang, Zhuoming and Fan, Gaofeng and Jia, Junji",
    title = "{Effect of particle spin on trajectory deflection and gravitational lensing}",
    eprint = "2207.09194",
    archivePrefix = "arXiv",
    primaryClass = "gr-qc",
    doi = "10.1088/1475-7516/2022/09/061",
    journal = "JCAP",
    volume = "09",
    pages = "061",
    year = "2022"
}

@article{Bini:2017ldh,
    author = "Bini, Donato and Geralico, Andrea",
    title = "{Hyperbolic-like elastic scattering of spinning particles by a Schwarzschild black hole}",
    eprint = "1808.06502",
    archivePrefix = "arXiv",
    primaryClass = "gr-qc",
    doi = "10.1007/s10714-017-2247-2",
    journal = "Gen. Rel. Grav.",
    volume = "49",
    number = "6",
    pages = "84",
    year = "2017"
}

@article{Bini:2017pee,
    author = "Bini, Donato and Geralico, Andrea and Vines, Justin",
    title = "{Hyperbolic scattering of spinning particles by a Kerr black hole}",
    eprint = "1707.09814",
    archivePrefix = "arXiv",
    primaryClass = "gr-qc",
    doi = "10.1103/PhysRevD.96.084044",
    journal = "Phys. Rev. D",
    volume = "96",
    number = "8",
    pages = "084044",
    year = "2017"
}

@article{Perlick:2004tq,
    author = "Perlick, V.",
    title = "{Gravitational lensing from a spacetime perspective}",
    doi = "10.12942/lrr-2004-9",
    journal = "Living Rev. Rel.",
    volume = "7",
    pages = "9",
    year = "2004"
}

@article{Werner:2012rc,
    author = "Werner, M. C.",
    title = "{Gravitational lensing in the Kerr-Randers optical geometry}",
    eprint = "1205.3876",
    archivePrefix = "arXiv",
    primaryClass = "gr-qc",
    doi = "10.1007/s10714-012-1458-9",
    journal = "Gen. Rel. Grav.",
    volume = "44",
    pages = "3047--3057",
    year = "2012"
}

@article{Perlick_1990,
  author    = {Perlick, V},
  journal   = {Classical and Quantum Gravity},
  title     = {On Fermat’s principle in general relativity. I. The general case},
  year      = {1990},
  issn      = {1361-6382},
  month     = aug,
  number    = {8},
  pages     = {1319--1331},
  volume    = {7},
  doi       = {10.1088/0264-9381/7/8/011},
  publisher = {IOP Publishing},
}

@article{Takizawa:2020egm,
    author = "Takizawa, Keita and Ono, Toshiaki and Asada, Hideki",
    title = "{Gravitational deflection angle of light: Definition by an observer and its application to an asymptotically nonflat spacetime}",
    eprint = "2001.03290",
    archivePrefix = "arXiv",
    primaryClass = "gr-qc",
    doi = "10.1103/PhysRevD.101.104032",
    journal = "Phys. Rev. D",
    volume = "101",
    number = "10",
    pages = "104032",
    year = "2020"
}

@article{Ono:2019hkw,
    author = "Ono, Toshiaki and Asada, Hideki",
    title = "{The effects of finite distance on the gravitational deflection angle of light}",
    eprint = "1906.02414",
    archivePrefix = "arXiv",
    primaryClass = "gr-qc",
    doi = "10.3390/universe5110218",
    journal = "Universe",
    volume = "5",
    number = "11",
    pages = "218",
    year = "2019"
}

@article{Crisnejo:2018uyn,
    author = "Crisnejo, Gabriel and Gallo, Emanuel",
    title = "{Weak lensing in a plasma medium and gravitational deflection of massive particles using the Gauss-Bonnet theorem. A unified treatment}",
    eprint = "1804.05473",
    archivePrefix = "arXiv",
    primaryClass = "gr-qc",
    doi = "10.1103/PhysRevD.97.124016",
    journal = "Phys. Rev. D",
    volume = "97",
    number = "12",
    pages = "124016",
    year = "2018"
}

@article{Crisnejo:2019ril,
    author = "Crisnejo, Gabriel and Gallo, Emanuel and Jusufi, Kimet",
    title = "{Higher order corrections to deflection angle of massive particles and light rays in plasma media for stationary spacetimes using the Gauss-Bonnet theorem}",
    eprint = "1910.02030",
    archivePrefix = "arXiv",
    primaryClass = "gr-qc",
    doi = "10.1103/PhysRevD.100.104045",
    journal = "Phys. Rev. D",
    volume = "100",
    number = "10",
    pages = "104045",
    year = "2019"
}

@article{Wald:1972sz,
    author = "Wald, Robert M.",
    title = "{Gravitational spin interaction}",
    doi = "10.1103/PhysRevD.6.406",
    journal = "Phys. Rev. D",
    volume = "6",
    pages = "406--413",
    year = "1972"
}

@article{Hanson:1974qy,
    author = "Hanson, Andrew J. and Regge, T.",
    title = "{The Relativistic Spherical Top}",
    reportNumber = "Print-74-0833 (IAS,PRINCETON)",
    doi = "10.1016/0003-4916(74)90046-3",
    journal = "Annals Phys.",
    volume = "87",
    pages = "498",
    year = "1974"
}

@article{Bailey:1975fe,
    author = "Bailey, I. and Israel, W.",
    title = "{Lagrangian Dynamics of Spinning Particles and Polarized Media in General Relativity}",
    doi = "10.1007/BF01609434",
    journal = "Commun. Math. Phys.",
    volume = "42",
    pages = "65--82",
    year = "1975"
}

@article{Lukes-Gerakopoulos:2014dma,
    author = "Lukes-Gerakopoulos, Georgios and Seyrich, Jonathan and Kunst, Daniela",
    title = "{Investigating spinning test particles: spin supplementary conditions and the Hamiltonian formalism}",
    eprint = "1409.4314",
    archivePrefix = "arXiv",
    primaryClass = "gr-qc",
    doi = "10.1103/PhysRevD.90.104019",
    journal = "Phys. Rev. D",
    volume = "90",
    number = "10",
    pages = "104019",
    year = "2014"
}

@article{Ono:2017pie,
    author = "Ono, Toshiaki and Ishihara, Asahi and Asada, Hideki",
    title = "{Gravitomagnetic bending angle of light with finite-distance corrections in stationary axisymmetric spacetimes}",
    eprint = "1704.05615",
    archivePrefix = "arXiv",
    primaryClass = "gr-qc",
    doi = "10.1103/PhysRevD.96.104037",
    journal = "Phys. Rev. D",
    volume = "96",
    number = "10",
    pages = "104037",
    year = "2017"
}

@article{Li:2019mqw,
    author = "Li, Zonghai and Zhou, Tao",
    title = "{Equivalence of Gibbons-Werner method to geodesics method in the study of gravitational lensing}",
    eprint = "1908.05592",
    archivePrefix = "arXiv",
    primaryClass = "gr-qc",
    doi = "10.1103/PhysRevD.101.044043",
    journal = "Phys. Rev. D",
    volume = "101",
    number = "4",
    pages = "044043",
    year = "2020"
}

@article{Eiroa:2002mk,
    author = "Eiroa, Ernesto F. and Romero, Gustavo E. and Torres, Diego F.",
    title = "{Reissner-Nordstrom black hole lensing}",
    eprint = "gr-qc/0203049",
    archivePrefix = "arXiv",
    doi = "10.1103/PhysRevD.66.024010",
    journal = "Phys. Rev. D",
    volume = "66",
    pages = "024010",
    year = "2002"
}

@article{He:2020eah,
    author = "He, Guansheng and Zhou, Xia and Feng, Zhongwen and Mu, Xueling and Wang, Hui and Li, Weijun and Pan, Chaohong and Lin, Wenbin",
    title = "{Gravitational deflection of massive particles in Schwarzschild-de Sitter spacetime}",
    doi = "10.1140/epjc/s10052-020-8382-z",
    journal = "Eur. Phys. J. C",
    volume = "80",
    number = "9",
    pages = "835",
    year = "2020"
}

@article{Arakida:2011ty,
    author = "Arakida, Hideyoshi and Kasai, Masumi",
    title = "{Effect of the cosmological constant on the bending of light and the cosmological lens equation}",
    eprint = "1110.6735",
    archivePrefix = "arXiv",
    primaryClass = "gr-qc",
    doi = "10.1103/PhysRevD.85.023006",
    journal = "Phys. Rev. D",
    volume = "85",
    pages = "023006",
    year = "2012"
}

@article{Rindler:2007zz,
    author = "Rindler, Wolfgang and Ishak, Mustapha",
    title = "{Contribution of the cosmological constant to the relativistic bending of light revisited}",
    eprint = "0709.2948",
    archivePrefix = "arXiv",
    primaryClass = "astro-ph",
    doi = "10.1103/PhysRevD.76.043006",
    journal = "Phys. Rev. D",
    volume = "76",
    pages = "043006",
    year = "2007"
}

@article{Sereno:2007rm,
    author = "Sereno, M.",
    title = "{On the influence of the cosmological constant on gravitational lensing in small systems}",
    eprint = "0711.1802",
    archivePrefix = "arXiv",
    primaryClass = "astro-ph",
    doi = "10.1103/PhysRevD.77.043004",
    journal = "Phys. Rev. D",
    volume = "77",
    pages = "043004",
    year = "2008"
}

@article{Vines:2018gqi,
    author = "Vines, Justin and Steinhoff, Jan and Buonanno, Alessandra",
    title = "{Spinning-black-hole scattering and the test-black-hole limit at second post-Minkowskian order}",
    eprint = "1812.00956",
    archivePrefix = "arXiv",
    primaryClass = "gr-qc",
    doi = "10.1103/PhysRevD.99.064054",
    journal = "Phys. Rev. D",
    volume = "99",
    number = "6",
    pages = "064054",
    year = "2019"
}

@article{Guevara:2018wpp,
    author = "Guevara, Alfredo and Ochirov, Alexander and Vines, Justin",
    title = "{Scattering of Spinning Black Holes from Exponentiated Soft Factors}",
    eprint = "1812.06895",
    archivePrefix = "arXiv",
    primaryClass = "hep-th",
    doi = "10.1007/JHEP09(2019)056",
    journal = "JHEP",
    volume = "09",
    pages = "056",
    year = "2019"
}

@article{Ben-Shahar:2023djm,
    author = "Ben-Shahar, Maor",
    title = "{Scattering of spinning compact objects from a worldline EFT}",
    eprint = "2311.01430",
    archivePrefix = "arXiv",
    primaryClass = "hep-th",
    reportNumber = "UUITP-31/23",
    doi = "10.1007/JHEP03(2024)108",
    journal = "JHEP",
    volume = "03",
    pages = "108",
    year = "2024"
}

@article{Li:2019vhp,
    author = "Li, Zonghai and He, Guansheng and Zhou, Tao",
    title = "{Gravitational deflection of relativistic massive particles by wormholes}",
    eprint = "1908.01647",
    archivePrefix = "arXiv",
    primaryClass = "gr-qc",
    doi = "10.1103/PhysRevD.101.044001",
    journal = "Phys. Rev. D",
    volume = "101",
    number = "4",
    pages = "044001",
    year = "2020"
}

@article{Abdujabbarov:2017pfw,
    author = "Abdujabbarov, Ahmadjon and Ahmedov, Bobomurat and Dadhich, Naresh and Atamurotov, Farruh",
    title = "{Optical properties of a braneworld black hole: Gravitational lensing and retrolensing}",
    doi = "10.1103/PhysRevD.96.084017",
    journal = "Phys. Rev. D",
    volume = "96",
    number = "8",
    pages = "084017",
    year = "2017"
}

@article{Morozova:2013uyv,
    author = "Morozova, V. S. and Ahmedov, B. J. and Tursunov, A. A.",
    title = "{Gravitational lensing by a rotating massive object in a plasma}",
    doi = "10.1007/s10509-013-1458-6",
    journal = "Astrophys. Space Sci.",
    volume = "346",
    number = "2",
    pages = "513--520",
    year = "2013"
}

@article{Schee:2017hof,
    author = "Schee, Jan and Stuchl{\'\i}k, Zden{\v{e}}k and Ahmedov, Bobomurat and Abdujabbarov, Ahmadjon and Toshmatov, Bobir",
    editor = "De Paolis, Francesco and Gurzadyan, Vahe G. and Siddiqui, Azad A. and Nucita, Achille",
    title = "{Gravitational lensing by regular black holes surrounded by plasma}",
    doi = "10.1142/S0218271817410115",
    journal = "Int. J. Mod. Phys. D",
    volume = "26",
    number = "5",
    pages = "1741011",
    year = "2017"
}

@article{Turimov:2018ttf,
    author = "Turimov, Bobur and Ahmedov, Bobomurat and Abdujabbarov, Ahmadjon and Bambi, Cosimo",
    title = "{Gravitational lensing by a magnetized compact object in the presence of plasma}",
    eprint = "1802.03293",
    archivePrefix = "arXiv",
    primaryClass = "gr-qc",
    doi = "10.1142/S0218271820400131",
    journal = "Int. J. Mod. Phys. D",
    volume = "28",
    number = "16",
    pages = "2040013",
    year = "2019"
}

@article{Wang:2025fmz,
    author = "Wang, Zi-Liang and Battista, Emmanuele",
    title = "{Dynamical features and shadows of quantum Schwarzschild black hole in effective field theories of gravity}",
    eprint = "2501.14516",
    archivePrefix = "arXiv",
    primaryClass = "gr-qc",
    doi = "10.1140/epjc/s10052-025-13833-7",
    journal = "Eur. Phys. J. C",
    volume = "85",
    number = "3",
    pages = "304",
    year = "2025"
}

@article{Capozziello:2025wwl,
    author = "Capozziello, Salvatore and Battista, Emmanuele and De Bianchi, Silvia",
    title = "{Null geodesics, causal structure, and matter accretion in Lorentzian-Euclidean black holes}",
    eprint = "2507.08431",
    archivePrefix = "arXiv",
    primaryClass = "gr-qc",
    doi = "10.1103/ybjp-8w2w",
    journal = "Phys. Rev. D",
    volume = "112",
    number = "4",
    pages = "044009",
    year = "2025"
}

@article{DeBianchi:2025bgn,
    author = "De Bianchi, Silvia and Capozziello, Salvatore and Battista, Emmanuele",
    title = "{Atemporality from Conservation Laws of Physics in Lorentzian-Euclidean Black Holes}",
    eprint = "2504.17570",
    archivePrefix = "arXiv",
    primaryClass = "gr-qc",
    doi = "10.1007/s10701-025-00848-z",
    journal = "Found. Phys.",
    volume = "55",
    number = "3",
    pages = "36",
    year = "2025"
}

@article{Capozziello:2024ucm,
    author = "Capozziello, Salvatore and De Bianchi, Silvia and Battista, Emmanuele",
    title = "{Avoiding singularities in Lorentzian-Euclidean black holes: The role of~atemporality}",
    eprint = "2404.17267",
    archivePrefix = "arXiv",
    primaryClass = "gr-qc",
    doi = "10.1103/PhysRevD.109.104060",
    journal = "Phys. Rev. D",
    volume = "109",
    number = "10",
    pages = "104060",
    year = "2024"
}

@article{Khodadi:2025wuw,
    author = "Khodadi, Mohsen and Lambiase, Gaetano and Mastrototaro, Leonardo and Poddar, Tanmay Kumar",
    title = "{Primordial gravitational waves from spontaneous Lorentz symmetry breaking}",
    eprint = "2501.14395",
    archivePrefix = "arXiv",
    primaryClass = "astro-ph.CO",
    doi = "10.1016/j.physletb.2025.139597",
    journal = "Phys. Lett. B",
    volume = "867",
    pages = "139597",
    year = "2025"
}

@article{Pantig:2022gih,
    author = {Pantig, Reggie C. and Mastrototaro, Leonardo and Lambiase, Gaetano and {\"O}vg{\"u}n, Ali},
    title = "{Shadow, lensing, quasinormal modes, greybody bounds and neutrino propagation by dyonic ModMax black holes}",
    eprint = "2208.06664",
    archivePrefix = "arXiv",
    primaryClass = "gr-qc",
    doi = "10.1140/epjc/s10052-022-11125-y",
    journal = "Eur. Phys. J. C",
    volume = "82",
    number = "12",
    pages = "1155",
    year = "2022"
}

@article{Zhong:2024ysg,
    author = "Zhong, Zhen and Cardoso, Vitor and Chen, Yifan",
    title = "{Dynamical Lensing Tomography of Black Hole Ringdowns}",
    eprint = "2408.10303",
    archivePrefix = "arXiv",
    primaryClass = "gr-qc",
    doi = "10.1103/PhysRevLett.134.211402",
    journal = "Phys. Rev. Lett.",
    volume = "134",
    number = "21",
    pages = "211402",
    year = "2025"
}

@article{Alloqulov:2024sns,
    author = "Alloqulov, Mirzabek and Chakrabarty, Hrishikesh and Malafarina, Daniele and Ahmedov, Bobomurat and Abdujabbarov, Ahmadjon",
    title = "{Gravitational lensing of neutrinos in parametrized black hole spacetimes}",
    eprint = "2408.12916",
    archivePrefix = "arXiv",
    primaryClass = "gr-qc",
    doi = "10.1088/1475-7516/2025/02/070",
    journal = "JCAP",
    volume = "02",
    pages = "070",
    year = "2025"
}

@article{Turakhonov:2024xfg,
    author = "Turakhonov, Ziyodulla and Hoshimov, Husanboy and Atamurotov, Farruh and Ghosh, Sushant G. and Abdujabbarov, Ahmadjon",
    title = "{Observational signatures of strong gravitational lensing in GUP-modified Schwarzschild black holes}",
    doi = "10.1016/j.dark.2024.101716",
    journal = "Phys. Dark Univ.",
    volume = "46",
    pages = "101716",
    year = "2024"
}

@article{Ditta:2025vsa,
    author = "Ditta, Allah and Bouzenada, Abdelmalek and Mustafa, G. and Javed, Faisal and Afandi, Fakhranda and Mahmood, Asif",
    title = "{Impact of gravitational collapse exhibiting loop quantum black holes on thermodynamical features and weak gravitational lensing}",
    doi = "10.1016/j.dark.2025.101818",
    journal = "Phys. Dark Univ.",
    volume = "47",
    pages = "101818",
    year = "2025"
}

@article{Molla:2025yoh,
    author = "Molla, Niyaz Uddin and Chaudhary, Himanshu and Capozziello, Salvatore and Atamurotov, Farruh and Mustafa, G. and Debnath, Ujjal",
    title = "{Observable signatures of RN black holes with dark matter halos via strong gravitational lensing and constraints from EHT observations}",
    eprint = "2501.09439",
    archivePrefix = "arXiv",
    primaryClass = "gr-qc",
    doi = "10.1016/j.dark.2024.101804",
    journal = "Phys. Dark Univ.",
    volume = "47",
    pages = "101804",
    year = "2025"
}

@article{Vagnozzi:2022moj,
    author = "Vagnozzi, Sunny and others",
    title = "{Horizon-scale tests of gravity theories and fundamental physics from the Event Horizon Telescope image of Sagittarius A}",
    eprint = "2205.07787",
    archivePrefix = "arXiv",
    primaryClass = "gr-qc",
    reportNumber = "UCI-HEP-TR-2022-07",
    doi = "10.1088/1361-6382/acd97b",
    journal = "Class. Quant. Grav.",
    volume = "40",
    number = "16",
    pages = "165007",
    year = "2023"
}

@article{Khodadi:2020jij,
    author = "Khodadi, Mohsen and Allahyari, Alireza and Vagnozzi, Sunny and Mota, David F.",
    title = "{Black holes with scalar hair in light of the Event Horizon Telescope}",
    eprint = "2005.05992",
    archivePrefix = "arXiv",
    primaryClass = "gr-qc",
    doi = "10.1088/1475-7516/2020/09/026",
    journal = "JCAP",
    volume = "09",
    pages = "026",
    year = "2020"
}

@article{Allahyari:2019jqz,
    author = "Allahyari, Alireza and Khodadi, Mohsen and Vagnozzi, Sunny and Mota, David F.",
    title = "{Magnetically charged black holes from non-linear electrodynamics and the Event Horizon Telescope}",
    eprint = "1912.08231",
    archivePrefix = "arXiv",
    primaryClass = "gr-qc",
    doi = "10.1088/1475-7516/2020/02/003",
    journal = "JCAP",
    volume = "02",
    pages = "003",
    year = "2020"
}

@article{Kaiser:1996wk,
    author = "Kaiser, Nick and Jaffe, Andrew H.",
    title = "{Bending of light by gravity waves}",
    eprint = "astro-ph/9609043",
    archivePrefix = "arXiv",
    reportNumber = "CITA-96-13",
    doi = "10.1086/304357",
    journal = "Astrophys. J.",
    volume = "484",
    pages = "545--554",
    year = "1997"
}

@article{Damour:1998jm,
    author = "Damour, Thibault and Esposito-Farese, Gilles",
    title = "{Light deflection by gravitational waves from localized sources}",
    eprint = "gr-qc/9802019",
    archivePrefix = "arXiv",
    reportNumber = "IHES-P-98-11, CPT-98-P-3607",
    doi = "10.1103/PhysRevD.58.044003",
    journal = "Phys. Rev. D",
    volume = "58",
    pages = "044003",
    year = "1998"
}

@article{Tsukamoto:2016jzh,
    author = "Tsukamoto, Naoki",
    title = "{Deflection angle in the strong deflection limit in a general asymptotically flat, static, spherically symmetric spacetime}",
    eprint = "1612.08251",
    archivePrefix = "arXiv",
    primaryClass = "gr-qc",
    doi = "10.1103/PhysRevD.95.064035",
    journal = "Phys. Rev. D",
    volume = "95",
    number = "6",
    pages = "064035",
    year = "2017"
}

@article{Virbhadra:2024xpk,
    author = "Virbhadra, K. S.",
    title = "{Conservation of distortion of gravitationally lensed images}",
    eprint = "2402.17190",
    archivePrefix = "arXiv",
    primaryClass = "gr-qc",
    doi = "10.1103/PhysRevD.109.124004",
    journal = "Phys. Rev. D",
    volume = "109",
    number = "12",
    pages = "124004",
    year = "2024"
}

@article{Adler:2022qtb,
    author = "Adler, Stephen L. and Virbhadra, K. S.",
    title = "{Cosmological constant corrections to the photon sphere and black hole shadow radii}",
    eprint = "2205.04628",
    archivePrefix = "arXiv",
    primaryClass = "gr-qc",
    doi = "10.1007/s10714-022-02976-7",
    journal = "Gen. Rel. Grav.",
    volume = "54",
    number = "8",
    pages = "93",
    year = "2022"
}

@article{Virbhadra:2022ybp,
    author = "Virbhadra, K. S.",
    title = "{Compactness of supermassive dark objects at galactic centers}",
    eprint = "2204.01792",
    archivePrefix = "arXiv",
    primaryClass = "gr-qc",
    doi = "10.1139/cjp-2023-0313",
    journal = "Can. J. Phys.",
    volume = "102",
    pages = "512",
    year = "2024"
}

@article{Virbhadra:2022iiy,
    author = "Virbhadra, K. S.",
    title = "{Distortions of images of Schwarzschild lensing}",
    eprint = "2204.01879",
    archivePrefix = "arXiv",
    primaryClass = "gr-qc",
    doi = "10.1103/PhysRevD.106.064038",
    journal = "Phys. Rev. D",
    volume = "106",
    number = "6",
    pages = "064038",
    year = "2022"
}

@article{Virbhadra:2008ws,
    author = "Virbhadra, K. S.",
    title = "{Relativistic images of Schwarzschild black hole lensing}",
    eprint = "0810.2109",
    archivePrefix = "arXiv",
    primaryClass = "gr-qc",
    doi = "10.1103/PhysRevD.79.083004",
    journal = "Phys. Rev. D",
    volume = "79",
    pages = "083004",
    year = "2009"
}

@article{Virbhadra:2007kw,
    author = "Virbhadra, K. S. and Keeton, C. R.",
    title = "{Time delay and magnification centroid due to gravitational lensing by black holes and naked singularities}",
    eprint = "0710.2333",
    archivePrefix = "arXiv",
    primaryClass = "gr-qc",
    doi = "10.1103/PhysRevD.77.124014",
    journal = "Phys. Rev. D",
    volume = "77",
    pages = "124014",
    year = "2008"
}

@article{Virbhadra:2002ju,
    author = "Virbhadra, K. S. and Ellis, G. F. R.",
    title = "{Gravitational lensing by naked singularities}",
    doi = "10.1103/PhysRevD.65.103004",
    journal = "Phys. Rev. D",
    volume = "65",
    pages = "103004",
    year = "2002"
}

@article{Claudel:2000yi,
    author = "Claudel, Clarissa-Marie and Virbhadra, K. S. and Ellis, G. F. R.",
    title = "{The Geometry of photon surfaces}",
    eprint = "gr-qc/0005050",
    archivePrefix = "arXiv",
    doi = "10.1063/1.1308507",
    journal = "J. Math. Phys.",
    volume = "42",
    pages = "818--838",
    year = "2001"
}

@article{Virbhadra:1999nm,
    author = "Virbhadra, K. S. and Ellis, George F. R.",
    title = "{Schwarzschild black hole lensing}",
    eprint = "astro-ph/9904193",
    archivePrefix = "arXiv",
    doi = "10.1103/PhysRevD.62.084003",
    journal = "Phys. Rev. D",
    volume = "62",
    pages = "084003",
    year = "2000"
}

@article{Virbhadra:1998kd,
    author = "Virbhadra, K. S.",
    title = "{Naked singularities and Seifert's conjecture}",
    eprint = "gr-qc/9809077",
    archivePrefix = "arXiv",
    doi = "10.1103/PhysRevD.60.104041",
    journal = "Phys. Rev. D",
    volume = "60",
    pages = "104041",
    year = "1999"
}

@article{Virbhadra:1998dy,
    author = "Virbhadra, K. S. and Narasimha, D. and Chitre, S. M.",
    title = "{Role of the scalar field in gravitational lensing}",
    eprint = "astro-ph/9801174",
    archivePrefix = "arXiv",
    journal = "Astron. Astrophys.",
    volume = "337",
    pages = "1--8",
    year = "1998"}

@article{Kopeikin:1999ev,
    author = "Kopeikin, Sergei M. and Schaefer, Gerhard",
    title = "{Lorentz covariant theory of light propagation in gravitational fields of arbitrary moving bodies}",
    eprint = "gr-qc/9902030",
    archivePrefix = "arXiv",
    doi = "10.1103/PhysRevD.60.124002",
    journal = "Phys. Rev. D",
    volume = "60",
    pages = "124002",
    year = "1999"
}

@article{Chen:2022kzv,
    author = "Chen, Yifan and Xue, Xiao and Brito, Richard and Cardoso, Vitor",
    title = "{Photon Ring Astrometry for Superradiant Clouds}",
    eprint = "2211.03794",
    archivePrefix = "arXiv",
    primaryClass = "gr-qc",
    reportNumber = "DESY 22-170",
    doi = "10.1103/PhysRevLett.130.111401",
    journal = "Phys. Rev. Lett.",
    volume = "130",
    number = "11",
    pages = "111401",
    year = "2023"
}

@article{Johnson:2019ljv,
    author = "Johnson, Michael D. and others",
    title = "{Universal interferometric signatures of a black hole{\textquoteright}s photon ring}",
    eprint = "1907.04329",
    archivePrefix = "arXiv",
    primaryClass = "astro-ph.IM",
    doi = "10.1126/sciadv.aaz1310",
    journal = "Sci. Adv.",
    volume = "6",
    number = "12",
    pages = "eaaz1310",
    year = "2020"
}

@article{Schwarzschild:1916uq,
    author = "Schwarzschild, Karl",
    title = "{On the gravitational field of a mass point according to Einstein's theory}",
    eprint = "physics/9905030",
    archivePrefix = "arXiv",
    journal = "Sitzungsber. Preuss. Akad. Wiss. Berlin (Math. Phys. )",
    volume = "1916",
    pages = "189--196",
    year = "1916"
}

@book{Chandrasekhar:1985kt,
    author = "Chandrasekhar, Subrahmanyan",
    title = "{The mathematical theory of black holes}",
    isbn = "978-0-19-850370-5",
    publisher = "Oxford University Press",
    year = "1985"
}

@article{Regge:1957td,
    author = "Regge, Tullio and Wheeler, John A.",
    title = "{Stability of a Schwarzschild singularity}",
    doi = "10.1103/PhysRev.108.1063",
    journal = "Phys. Rev.",
    volume = "108",
    pages = "1063--1069",
    year = "1957"
}

@article{Zerilli:1970wzz,
    author = "Zerilli, F. J.",
    title = "{Gravitational field of a particle falling in a schwarzschild geometry analyzed in tensor harmonics}",
    doi = "10.1103/PhysRevD.2.2141",
    journal = "Phys. Rev. D",
    volume = "2",
    pages = "2141--2160",
    year = "1970"
}

@article{Martel:2005ir,
    author = "Martel, Karl and Poisson, Eric",
    title = "{Gravitational perturbations of the Schwarzschild spacetime: A Practical covariant and gauge-invariant formalism}",
    eprint = "gr-qc/0502028",
    archivePrefix = "arXiv",
    doi = "10.1103/PhysRevD.71.104003",
    journal = "Phys. Rev. D",
    volume = "71",
    pages = "104003",
    year = "2005"
}

@article{Moncrief:1974am,
    author = "Moncrief, V.",
    title = "{Gravitational perturbations of spherically symmetric systems. I. The exterior problem.}",
    doi = "10.1016/0003-4916(74)90173-0",
    journal = "Annals Phys.",
    volume = "88",
    pages = "323--342",
    year = "1974"
}

@article{Berti:2009kk,
    author = "Berti, Emanuele and Cardoso, Vitor and Starinets, Andrei O.",
    title = "{Quasinormal modes of black holes and black branes}",
    eprint = "0905.2975",
    archivePrefix = "arXiv",
    primaryClass = "gr-qc",
    doi = "10.1088/0264-9381/26/16/163001",
    journal = "Class. Quant. Grav.",
    volume = "26",
    pages = "163001",
    year = "2009"
}

@Article{Cunningham_1972,
  author    = {Cunningham, C. T. and Bardeen, J. M.},
  journal   = {The Astrophysical Journal},
  title     = {The Optical Appearance of a Star Orbiting an Extreme Kerr Black Hole},
  year      = {1972},
  issn      = {1538-4357},
  month     = may,
  pages     = {L137},
  volume    = {173},
  doi       = {10.1086/180933},
  publisher = {American Astronomical Society},
}

@book{Weinberg:1972kfs,
    author = "Weinberg, Steven",
    title = "{Gravitation and Cosmology}: {Principles and Applications of the General Theory of Relativity}",
    isbn = "978-0-471-92567-5, 978-0-471-92567-5",
    publisher = "John Wiley and Sons",
    address = "New York",
    year = "1972"
}

@Article{Einstein_2005,
  author    = {Einstein, A.},
  journal   = {Albert Einstein: Akademie‐Vorträge},
  title     = {Erklärung der Perihelbewegung des Merkur aus der allgemeinen Relativitätstheorie},
  year      = {2005},
  month     = dec,
  pages     = {78--87},
  doi       = {10.1002/3527608958.ch4},
  isbn      = {9783527608959},
  publisher = {Wiley},
}

@article{Bozza:2002zj,
    author = "Bozza, V.",
    title = "{Gravitational lensing in the strong field limit}",
    eprint = "gr-qc/0208075",
    archivePrefix = "arXiv",
    doi = "10.1103/PhysRevD.66.103001",
    journal = "Phys. Rev. D",
    volume = "66",
    pages = "103001",
    year = "2002"
}

@article{Will:2014kxa,
    author = "Will, Clifford M.",
    title = "{The Confrontation between General Relativity and Experiment}",
    eprint = "1403.7377",
    archivePrefix = "arXiv",
    primaryClass = "gr-qc",
    doi = "10.12942/lrr-2014-4",
    journal = "Living Rev. Rel.",
    volume = "17",
    pages = "4",
    year = "2014"
}

@article{Perlick:2021aok,
    author = "Perlick, Volker and Tsupko, Oleg Yu.",
    title = "{Calculating black hole shadows: Review of analytical studies}",
    eprint = "2105.07101",
    archivePrefix = "arXiv",
    primaryClass = "gr-qc",
    doi = "10.1016/j.physrep.2021.10.004",
    journal = "Phys. Rept.",
    volume = "947",
    pages = "1--39",
    year = "2022"
}

@article{Kokkotas:1999bd,
    author = "Kokkotas, Kostas D. and Schmidt, Bernd G.",
    title = "{Quasinormal modes of stars and black holes}",
    eprint = "gr-qc/9909058",
    archivePrefix = "arXiv",
    doi = "10.12942/lrr-1999-2",
    journal = "Living Rev. Rel.",
    volume = "2",
    pages = "2",
    year = "1999"
}

@article{Lewis:2006fu,
    author = "Lewis, Antony and Challinor, Anthony",
    title = "{Weak gravitational lensing of the CMB}",
    eprint = "astro-ph/0601594",
    archivePrefix = "arXiv",
    doi = "10.1016/j.physrep.2006.03.002",
    journal = "Phys. Rept.",
    volume = "429",
    pages = "1--65",
    year = "2006"
}

@article{Price:1971fb,
    author = "Price, Richard H.",
    title = "{Nonspherical perturbations of relativistic gravitational collapse. 1. Scalar and gravitational perturbations}",
    doi = "10.1103/PhysRevD.5.2419",
    journal = "Phys. Rev. D",
    volume = "5",
    pages = "2419--2438",
    year = "1972"
}

@article{Aratore:2024bro,
    author = "Aratore, Fabio and Tsupko, Oleg Yu. and Perlick, Volker",
    title = "{Constraining spherically symmetric metrics by the gap between photon rings}",
    eprint = "2402.14733",
    archivePrefix = "arXiv",
    primaryClass = "gr-qc",
    doi = "10.1103/PhysRevD.109.124057",
    journal = "Phys. Rev. D",
    volume = "109",
    number = "12",
    pages = "124057",
    year = "2024"
}

@article{Vincent:2022fwj,
    author = "Vincent, Frederic H. and Gralla, Samuel E. and Lupsasca, Alexandru and Wielgus, Maciek",
    title = "{Images and photon ring signatures of thick disks around black holes}",
    eprint = "2206.12066",
    archivePrefix = "arXiv",
    primaryClass = "astro-ph.HE",
    doi = "10.1051/0004-6361/202244339",
    journal = "Astron. Astrophys.",
    volume = "667",
    pages = "A170",
    year = "2022"
}

@article{Gralla:2019xty,
    author = "Gralla, Samuel E. and Holz, Daniel E. and Wald, Robert M.",
    title = "{Black Hole Shadows, Photon Rings, and Lensing Rings}",
    eprint = "1906.00873",
    archivePrefix = "arXiv",
    primaryClass = "astro-ph.HE",
    doi = "10.1103/PhysRevD.100.024018",
    journal = "Phys. Rev. D",
    volume = "100",
    number = "2",
    pages = "024018",
    year = "2019"
}

@misc{Dai:2017huk,
    author = "Dai, Liang and Venumadhav, Tejaswi",
    title = "{On the waveforms of gravitationally lensed gravitational waves}",
    eprint = "1702.04724",
    archivePrefix = "arXiv",
    primaryClass = "gr-qc",
    month = "2",
    year = "2017"
}

@article{EventHorizonTelescope:2019dse,
    author = "Akiyama, Kazunori and others",
    collaboration = "Event Horizon Telescope",
    title = "{First M87 Event Horizon Telescope Results. I. The Shadow of the Supermassive Black Hole}",
    eprint = "1906.11238",
    archivePrefix = "arXiv",
    primaryClass = "astro-ph.GA",
    doi = "10.3847/2041-8213/ab0ec7",
    journal = "Astrophys. J. Lett.",
    volume = "875",
    pages = "L1",
    year = "2019"
}

@article{EventHorizonTelescope:2022wkp,
    author = "Akiyama, Kazunori and others",
    collaboration = "Event Horizon Telescope",
    title = "{First Sagittarius A* Event Horizon Telescope Results. I. The Shadow of the Supermassive Black Hole in the Center of the Milky Way}",
    eprint = "2311.08680",
    archivePrefix = "arXiv",
    primaryClass = "astro-ph.HE",
    doi = "10.3847/2041-8213/ac6674",
    journal = "Astrophys. J. Lett.",
    volume = "930",
    number = "2",
    pages = "L12",
    year = "2022"
}

@article{Gralla:2020srx,
    author = "Gralla, Samuel E. and Lupsasca, Alexandru and Marrone, Daniel P.",
    title = "{The shape of the black hole photon ring: A precise test of strong-field general relativity}",
    eprint = "2008.03879",
    archivePrefix = "arXiv",
    primaryClass = "gr-qc",
    doi = "10.1103/PhysRevD.102.124004",
    journal = "Phys. Rev. D",
    volume = "102",
    number = "12",
    pages = "124004",
    year = "2020"
}

@article{Pantig:2024kfn,
    author = {Pantig, Reggie C. and {\"O}vg{\"u}n, Ali},
    title = "{Imprints of a gravitational wave through the weak field deflection of photons*}",
    eprint = "2406.05782",
    archivePrefix = "arXiv",
    primaryClass = "gr-qc",
    doi = "10.1088/1674-1137/ad4e25",
    journal = "Chin. Phys. C",
    volume = "48",
    number = "8",
    pages = "085104",
    year = "2024"
}

@article{Wang:2019skw,
    author = "Wang, Mingzhi and Chen, Songbai and Jing, Jiliang",
    title = "{Effect of gravitational wave on shadow of a Schwarzschild black hole}",
    eprint = "1908.04527",
    archivePrefix = "arXiv",
    primaryClass = "gr-qc",
    doi = "10.1140/epjc/s10052-021-09287-2",
    journal = "Eur. Phys. J. C",
    volume = "81",
    number = "6",
    pages = "509",
    year = "2021"
}

@article{Obukhov:2010kn,
    author = "Obukhov, Yuri N. and Puetzfeld, Dirk",
    title = "{Dynamics of test bodies with spin in de Sitter spacetime}",
    eprint = "1010.1451",
    archivePrefix = "arXiv",
    primaryClass = "gr-qc",
    doi = "10.1103/PhysRevD.83.044024",
    journal = "Phys. Rev. D",
    volume = "83",
    pages = "044024",
    year = "2011"
}

\end{document}